\documentclass[11pt,preprint]{aastex}
\usepackage{natbib}
\usepackage{graphicx}
\newcommand{\flux}{erg \,s$^{-1}$ \,cm$^{-2}$}

\newcommand{\km}{\,km\,s$^{-1}$}

\begin{document}

\title{Ultraviolet and X-ray Variability of the Seyfert 1.5 Galaxy Markarian 817}
\author{Lisa M. Winter\altaffilmark{1,2},  Charles Danforth\altaffilmark{1}, Ranjan Vasudevan\altaffilmark{3},W.\ N.\ Brandt\altaffilmark{3}, Jennifer Scott\altaffilmark{4}, Cynthia Froning\altaffilmark{1}, Brian Keeney\altaffilmark{1},
J. Michael Shull\altaffilmark{1}, Steve Penton\altaffilmark{1}, Richard Mushotzky\altaffilmark{5}, \\Donald P. Schneider\altaffilmark{3}, and Nahum Arav\altaffilmark{5}}
\altaffiltext{1}{Center for Astrophysics and Space Astronomy, Department of Astrophysical \& Planetary Sciences, University of Colorado, UCB 391, Boulder, CO 80309, USA}
\altaffiltext{2}{Hubble Fellow}
\altaffiltext{3}{Department of Astronomy and Astrophysics, The Pennsylvania State University, 525 Davey Laboratory, University Park, PA 16802, USA}
\altaffiltext{4}{Department of Physics, Astronomy, and Geosciences, Towson University, Towson, MD 21252, USA}
\altaffiltext{5}{Department of Astronomy, University of Maryland, College Park, MD 20742, USA}
\altaffiltext{6}{Department of Physics, Virginia Polytechnic Institute \& State University, Blacksburg, VA 24061, USA}
\begin{abstract}
We present an investigation of the ultraviolet and X-ray spectra of the Seyfert 1.5 galaxy Markarian 817.  The ultraviolet analysis includes two recent observations taken with the Cosmic Origins Spectrograph in August and December 2009, as well as archival spectra from the {\it International Ultraviolet Explorer} and the {\it Hubble Space Telescope}.  Twelve Ly$\alpha$ absorption features are detected in the 1997 GHRS and 2009 COS spectra -- of these, four are associated with high-velocity clouds in the interstellar medium, four are at low-significance, and the remaining four are intrinsic features, which vary between the GHRS and COS observations.  The strongest intrinsic absorber in the 1997 spectrum has a systemic velocity of $\sim -4250$\km.  The corresponding feature in the COS data is five times weaker than the GHRS absorber. The three additional weak (equivalent width from 13--54\,m\AA) intrinsic Ly$\alpha$ absorbers are at systemic velocities of $-4100$\km, $-3550$\km, and $-2600$\km.  However, intrinsic absorption troughs from highly ionized \ion{C}{4} and \ion{N}{5}, are not detected in the COS observations.  No ionized absorption signatures are detected in the $\sim 14$\,ks XMM-Newton EPIC spectra.  The factor of five change in the intrinsic Ly$\alpha$ absorber is most likely due to bulk motions in the absorber, since there is no drastic change in the UV luminosity of the source from the GHRS to the COS observations.  In a study of variability of Mrk 817, we find that the X-ray luminosity varies by a factor of $\sim 40$ over 20\,years, while the UV continuum/emission lines vary by at most a factor of $\sim 2.3$ over 30\,years.  The variability of the X-ray luminosity is strongly correlated with the X-ray power-law index, but no correlation is found with the simultaneous optical/UV photometry.
\end{abstract}
\keywords{galaxies: active, galaxies: Seyfert, galaxies: individual(Mrk 817), ultraviolet: galaxies, X-ray: galaxies}

\section{Introduction}
Long-term and often short-term variability in both flux and spectral shape, across the electromagnetic spectrum, is a fundamental characteristic of the spectra of active galactic nuclei (AGN).  While variability is ubiquitous in AGN, our understanding of long-term spectral changes in local AGNs is the result of studies of a few dozens of sources focusing primarily on variations in broad optical emission lines in conjunction with reverberation mapping studies (e.g.,~\citealt{2004ApJ...613..682P}) for local AGNs ($z < 0.1$). It is important to extend these studies in temporal space (e.g.,~acquiring a larger sample of observations over longer periods of time), the wavebands covered (e.g.,~multi-wavelength studies), and the number/type of sources sampled, in order to understand the physics behind the AGN emission -- the accretion processes, nature of outflows, and associated feedback.

In an effort to extend the variability studies of individual AGN, we present an analysis of the available UV and X-ray observations of Mrk 817, a nearby Seyfert 1.5 galaxy with a redshift of $z = 0.031158$ \citep{1999PASP..111..438F}.  Its optical spectrum exhibits complex emission in the narrow-line region (NLR) and broad-line region (BLR), with outflowing NLR [\ion{O}{3}] $\lambda\lambda 4959, 5007$\AA~emission measured at a systemic velocity of $-450$\,km\,s$^{-1}$ \citep{2006MNRAS.371.1610I}.  In the UV, observations with both the {\it Hubble Space Telescope} (HST) and the {\it Far Ultraviolet Spectroscopic Explorer} (FUSE) also revealed outflowing gas.  HST Goddard High-Resolution Spectrograph (GHRS) observations from 1997 of Ly$\alpha$ indicated the presence of an intrinsic AGN outflow \citep{2000ApJS..130..121P}.  Later observations with FUSE, in 2000--2001, show weak absorption in \ion{O}{6} with a radial systemic velocity component of $-$4100\,km\,s$^{-1}$, the largest offset measured in a Seyfert \citep{2007AJ....134.1061D}.  Analysis of multiple observations from FUSE shows that the weak absorption is variable in that the equivalent width of the absorber increases with decreasing continuum flux \citep{2008AJ....136.1201D}.  Additionally, a second and third absorption component ($v \approx -3000$\,\km) appeared in the final FUSE observation, potentially due to a change in ionization.

Owing to its UV brightness and sight line through a complex of Galactic high velocity clouds (HVC Complex C,
\citealt{2001AJ....122.3280G,2003ApJ...585..336C}),
Mrk 817 was a selected target for the guaranteed time observations with the HST Cosmic Origins Spectrograph (COS).  Two high signal-to-noise ratio observations were taken with COS, separated by $\sim 5$ months.  Additionally, UV observations of this source taken with the {\it International Ultraviolet Explorer} (IUE) and HST GHRS allow a search for variability over a period of nearly 30 years.  Details of the UV observations, are presented in Section~\ref{sect-uv}.  We include an analysis of the intrinsic AGN properties, including the UV continuum and emission-line properties from each of these observations.  In Section~\ref{sect-xray}, we present an analysis of X-ray spectra of Mrk 817, along with simultaneous optical/UV photometry,  from XMM-Newton and Swift.  Although the source is bright in the X-ray band,  it has only recently been targeted for observation in the X-ray with XMM-Newton.  Finally, we present our conclusions on the spectral analysis and UV and X-ray variability of Mrk 817 in Section~\ref{sect-conclusion}.  We assume a standard cosmology with $H_0 = 75$\km\,Mpc$^{-1}$, throughout.


\section{The Ultraviolet Spectra}~\label{sect-uv}
In this section, we include details of our analysis of the available UV spectra of Mrk 817.  In \S~\ref{sect-uvobservations}, we describe the observations from COS, GHRS, and IUE.  An analysis of the continuum and emission line properties from the COS observations is included in \S~\ref{sect-cos}.  In \S~\ref{sect-compareuv}, we compare the COS spectra with the GHRS and IUE spectra -- including detected Ly$\alpha$ absorbers in \S~\ref{sect-absorbers} and the continuum/emission-line properties in \S~\ref{sect-ghrsiue}.

\subsection{Observations}~\label{sect-uvobservations}


Mrk\,817 was observed twice with HST/COS with both the G130M (1135\,\AA\,$<\lambda<1480$\,\AA) and G160M (1400\,\AA\,$<\lambda<1795$\,\AA) medium-resolution ($R \equiv \lambda/\Delta \lambda\approx 18,000$) gratings.  The first epoch of observations (2009, August 4) were part of the Early Release Observations (ERO) program (PI: Noll), with four exposures in each grating for a total exposure time of 2162\,s (G130M) and 1410\,s (G160M).  The second set of exposures (2009, December 28) were taken as part of the COS Guaranteed Time Observations (GTO) program (PI: Green) and totaled 1264\,s and 1600\,s in G130M and G160M, respectively.  Both sets of observations used four different central wavelength settings per grating to dither instrumental features and provide continuous spectral coverage across the entire COS FUV band.  See \citet{2010AAS...21546418G}, \citet{2010AAS...21546422O}, and the COS HST Instrument Handbook \citep{COS-handbook} for details on the COS instrument and on-orbit performance.

The ERO data were obtained during the Servicing Mission Orbital Verification (SMOV) period before the instrument reached final focus.  However, the pre-focus data are indistinguishable from that taken after correct focus had been achieved, except in the case of the very narrowest of absorption features ($\Delta \lambda \la 0.1$\,\AA), none of which are analyzed in this work.  Another symptom unique to the early SMOV data was the ``divot and clod'' features which occurred near the blue end of each detector segment; charge resulting from photons hitting one part of the detector appeared in a different part.  The high voltage on the detectors was reduced on 2009, August 17, and the instrumental feature disappeared.  For the Mrk\,817 ERO observations, the divot and clod were carefully fitted with Gaussian profiles and normalized to the flux level in adjacent portions of the spectrum.  The wavelength region affected by this feature and its subsequent removal is relatively small, and much smaller in both amplitude and width than the AGN emission features analyzed here.  Thus, their effect on our scientific conclusions is minimal.

All data were reduced with {\sc CALCOS} v2.11f.  Partial flat-fielding, alignment, and coaddition of the processed exposures were carried out using IDL routines developed by the COS GTO team specifically for COS FUV data\footnote{IDL routines available at  {\tt \url{http://casa.colorado.edu/$\sim$danforth/costools.html}} and described in \citet{2010arXiv1005.2191D}.}.  Briefly, each exposure was corrected for narrow, $\sim15\%$-opaque shadows from ion repeller grid wires.  The local exposure time in these regions was reduced to give them less weight in the final co-addition.  Similarly, exposure times for data at the edges of the detectors was decreased to de-weight these data in the final co-addition.  With four different central wavelength settings per grating, any residual instrumental artifacts from grid-wire shadows and detector boundaries should have negligible effect on the final spectrum.

Next, strong interstellar features in each exposure were aligned via cross-correlation and interpolated onto a common wavelength scale.  The wavelength shifts were typically on the order of a resolution element ($\sim 0.07$~\AA, $\sim 17$\,\km) or less.  The coadded flux at each wavelength was taken to be the exposure-weighted mean of flux in each exposure.  

\begin{figure}
\centering
\includegraphics[width=15cm]{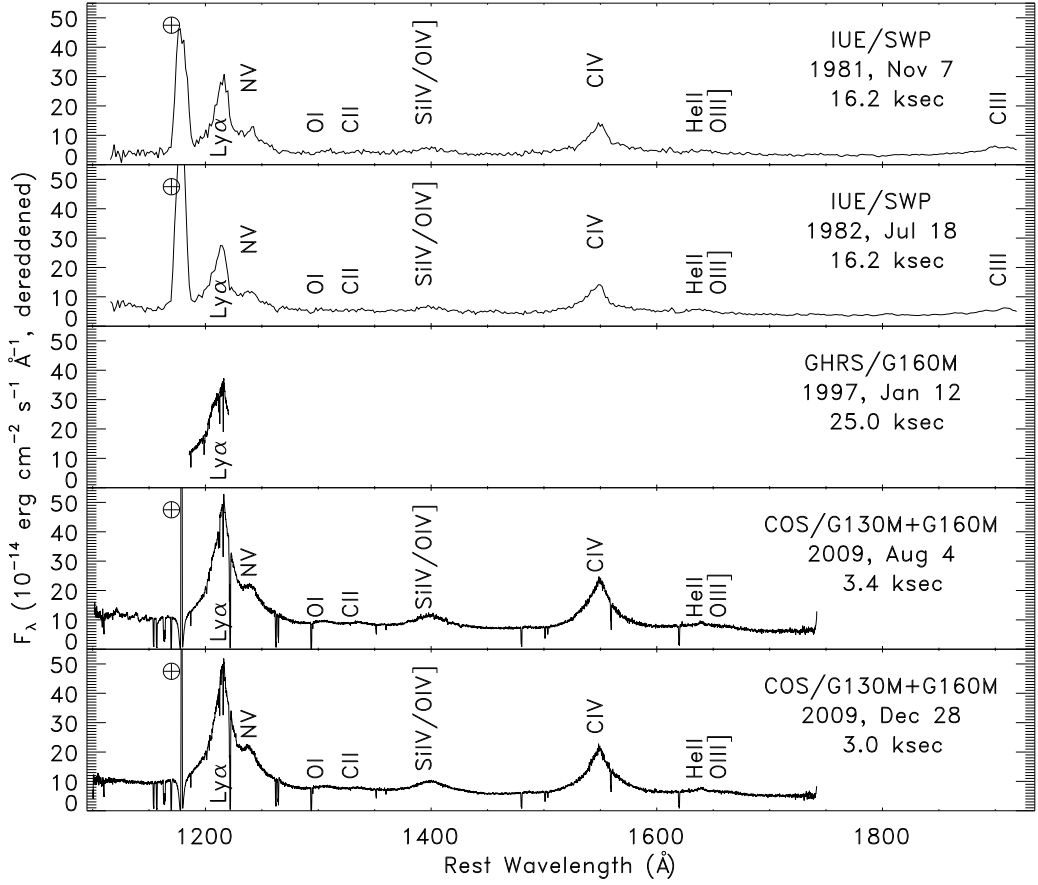}
\caption{Spectra of the UV observations from IUE, GHRS, and COS are shown, labeling
 the locations of emission lines frequently seen in AGN.  The spectra are corrected for Galactic extinction.  Additional absorption features in the spectrum are mostly associated with absorption within our galaxy.  The most prominent intrinsic AGN features are labelled.
\label{fig-UV}}
\end{figure}

The exposures from the ERO and GTO epochs were coadded separately and are presented in Figure~\ref{fig-UV}.  To quantify the quality of the combined data, we identify line-free continuum regions, smooth the data by the seven-pixel resolution element, and define $S/N\equiv \rm mean(flux)/stddev(flux)$.  In the 1150$-$1200 \AA\ band, we measure $S/N \approx37$\,per resolution element in the ERO and $S/N \approx26$ per resolution element in the GTO observation.  Similarly, in the 1500$-$1550\,\AA\ band, we measure $S/N \approx 32$ per resolution element in the ERO and $S/N \approx29$ per resolution element in the GTO observation.

In addition to the COS spectra, we also include archived IUE data from the Multimission Archive at STScI.  We downloaded large-aperture, low-dispersion grating observations of Mrk\,817 for two observations separated by approximately 8 months.  The observation IDs are SWP\,15447 and SWP\,17442 with observation dates of 1981 November 7 and 1982 July 18, respectively.  The exposure time for each of these observations is 16200\,s.  The archived data were downloaded in the finished processed form.  These spectra cover the AGN rest-frame wavelength range from $\sim 1120-2010$ \AA.

Finally, we include the HST GHRS spectrum of Mrk\,817.  The UV spectrum was observed on 1997 January 12 under project ID 6593 (PI: Stocke).  Three exposures were taken with the G160M grating for a total exposure time of 25\,ks.  The wavelength coverage of the GHRS spectrum is only $\approx 36$\AA, covering the intrinsic Ly$\alpha$ emission region.  We use the processed spectrum presented in \citet{2000ApJS..130..121P}, which includes the details of the calibration and spectral processing.

We summarize the details of all of the UV observations in Table~\ref{tbl-UV}, including the observation, observation date, exposure time, spectral coverage, and spectral resolution for the COS, GHRS, and IUE spectra.  Additionally, Figure~\ref{fig-UV} illustrates all of the UV spectra by order of their observation date.

\subsection{Analysis of the COS Spectra}\label{sect-cos}
For the analysis of each of the COS observations of Mrk 817, we first corrected the spectra for Galactic reddening, using the \citet{1990ARAA..28..215D} value of N$_{\rm H I} = 1.15 \times 10^{20}$\,cm$^{-2}$, which corresponds to $E(B-V) = 0.022$ following the relationship between hydrogen column density and interstellar reddening derived by \citet{1995AA...293..889P}.
We used this reddening value with the selective extinction curve of \citet{1989ApJ...345..245C} and the near-UV correction from \citet{1994ApJ...437..262O} and then shifted the spectra into the AGN rest-frame.  In Figure~\ref{fig-UV}, we show the resultant de-reddened  COS spectra (along with the de-reddened GHRS and IUE spectra).  The most prominent features are intrinsic AGN (and geo-coronal Ly$\alpha$) emission lines, and no strong intrinsic absorption is detected embedded in the broad emission line features.  There are, however, a number of weak absorption lines in Ly$\alpha$, which we will discuss in \S~\ref{sect-compareuv}.

We next chose continuum regions free of emission/absorption features to fit a power-law continuum of the form: $F_{\lambda}(\lambda) = a_0 \times (\lambda/1750$\,\AA$)^{-\alpha}$, where $a_0$ is the flux density in units of \flux\,\AA$^{-1}$~at 1750\,\AA~and $\alpha$ is the power-law index.  Our choice of 1750\AA~was arbitrary -- corresponding to a wavelength free of prominent emission/absorption at the end of the COS bandpass.  For our continuum regions, we chose 6 regions free of emission/absorption features from 1120--1700\AA, with individual region sizes from 10--50\AA.  The same continuum regions were used for both the ERO and GTO spectrum.  The resultant best-fit parameters, for both continuum and emission line fits, were determined using $\chi^2$-minimization along with the {\tt pyminuit}\footnote{\tt \url{http://code.google.com/p/pyminuit/}} python interface to the MINUIT -- Function Minimization and Error Analysis code\footnote{Copyright CERN, Geneva 1994-1998, with the MINUIT manual available at {\tt \url{http://wwwasdoc.web.cern.ch/wwwasdoc/minuit/minmain.html}}}~.  Results of these fits are shown in Table~\ref{tbl-powerlaw}, with recorded errors corresponding to the 1-$\sigma$ level.  We find that the power-law index is unchanged between the observations ($\alpha \approx 1.30 \pm 0.05$).  This slope is consistent with values obtained for other local AGN; for instance, the average slope found from an analysis of HST observations of narrow-line Seyfert 1s by \citet{2003PASP..115..592C} is $F_{\nu} \propto \nu^{-0.80}$ or $F_{\lambda} \propto \lambda^{-1.20}$ and \citet{2004ApJ...615..135S} find the composite EUV spectrum from FUSE of $z < 0.1$ sources as
$F_{\nu} \propto \nu^{-0.74}$ or $F_{\lambda} \propto \lambda^{-1.26}$.  We note that the slopes of higher redshift AGN tend to be different (i.e., \citet{2002ApJ...565..773T} find $F_{\nu} \propto \nu^{-1.76}$ or $F_{\lambda} \propto \lambda^{-0.24}$ from the HST spectra of 184 quasars with $z > 0.33$).  While the slope did not change appreciably between the COS ERO and GTO observations, the continuum level did vary, with the GTO observation being 30\% lower than the ERO observation.

The strongest features in the spectra of Mrk 817 are intrinsic AGN emission lines of Ly$\alpha$, the \ion{N}{5} doublet, the \ion{Si}{4} doublet (blended with \ion{O}{4}]), and the \ion{C}{4} doublet.  We fit the emission lines in these observations with a combination of Gaussians.  We chose Gaussians as the simplest mathematical representation of the lines and use them for a basic characterization of the emission-line properties.  Since the \ion{C}{4} doublet ($\lambda \lambda 1548.40, 1550.77$\AA) is the strongest feature that is not blended with nearby emission lines, we first fit \ion{C}{4} with a broad doublet and intermediate doublet component (four Gaussians).  Since the ratio of the statistical weights of the blue to red line is $2:1$, we fixed the flux of the blue line to be twice that of the flux in the red line.  We also fixed the separation of the lines in the doublet to the laboratory measured value of 2.6\,\AA.  Best-fit values for the \ion{C}{4} doublet, including the velocity offset, velocity width, equivalent width, and line flux, as well as for the other strong emission lines described below, are shown in Tables~\ref{tbl-emission-ero} and~\ref{tbl-emission-gto}. 

To fit the \ion{Si}{4} doublet ($\lambda \lambda 1393.76,1402.77$\,\AA), we fixed the velocity width of the line to that found in the \ion{C}{4} fits.  As for the \ion{C}{4} doublet fits, we fit the \ion{Si}{4} emission region with four Gaussians -- a broad and intermediate component for each line in the doublet -- fixing the difference between the line centers of each doublet component (broad or intermediate) to the laboratory measured difference between the lines in the doublet (9.015\AA) and the flux of the blue line to be twice that of the red line (the ratio of the line strengths is $2:1$).  We did not attempt to fit the \ion{Si}{4} and \ion{O}{4}] emission separately.  

Since Ly$\alpha$ ($\lambda 1215.67$\,\AA) and the \ion{N}{5} doublet ($\lambda \lambda 1238.82, 1242.80$) are blended, we fit these emission features simultaneously.  We first masked strong Galactic ISM lines in this region, including the absorption features from \ion{N}{5}, \ion{S}{2}, and \ion{Si}{2}.  We fit the \ion{N}{5} doublet with a broad and intermediate doublet composed of four Gaussians, fixing the width of the lines to be the same as that for the \ion{C}{4} doublet.  As for \ion{C}{4} and \ion{Si}{4}, we fixed the separation of the lines in the doublet to the laboratory measured value (3.980 \AA) and the blue line of the doublet to twice that of the red line (the ratio of the line strengths is also $2:1$ for \ion{N}{5}).  The asymmetric shape of the Ly$\alpha$ emission required us to fit this feature with a combination of three Gaussians.  We fixed the full width half-maximum for two of these Gaussians to the values obtained from the \ion{C}{4} fits.  

\begin{figure}
\centering
\includegraphics[width=8cm]{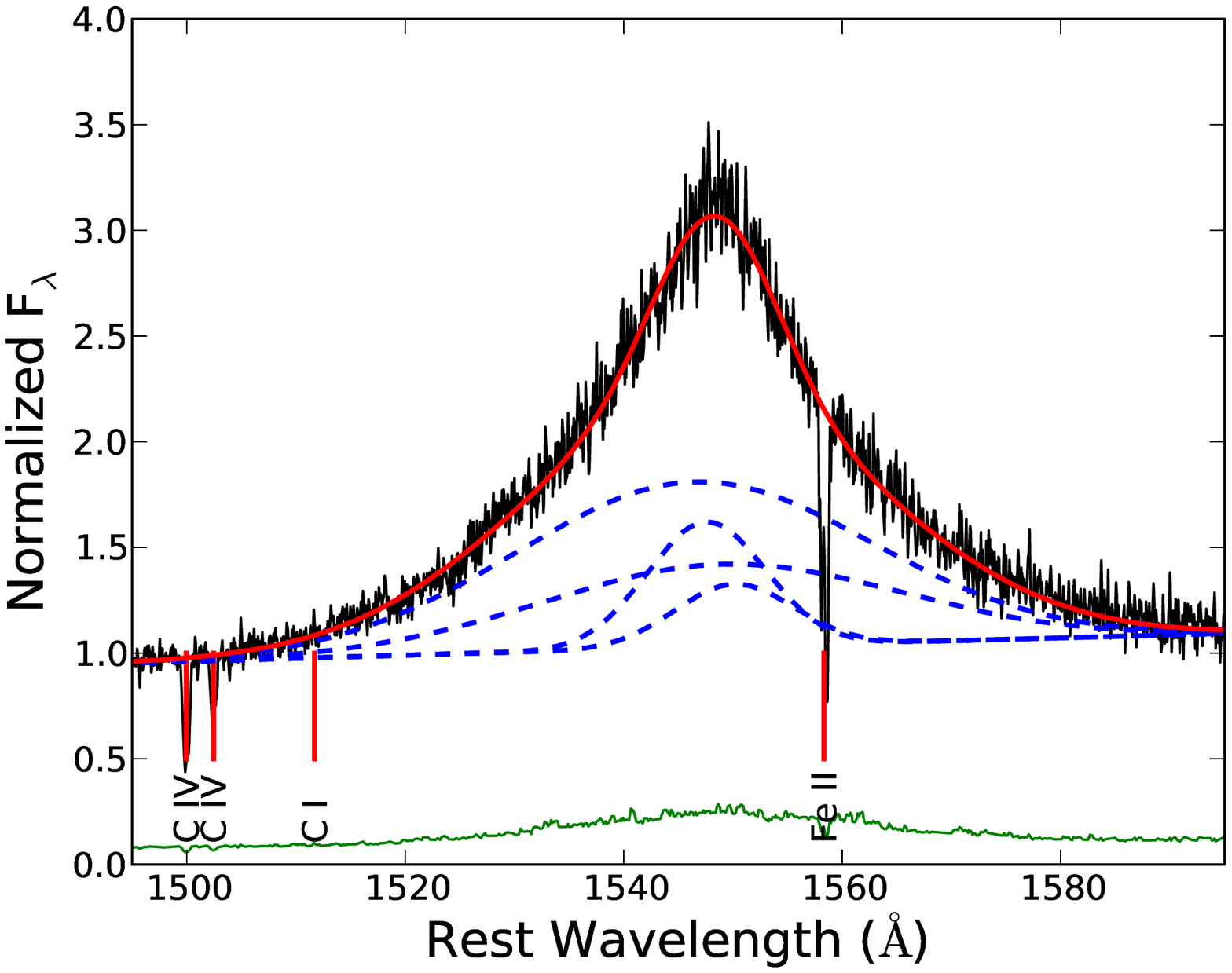} \\
\vspace{-0.07cm}
\includegraphics[width=8cm]{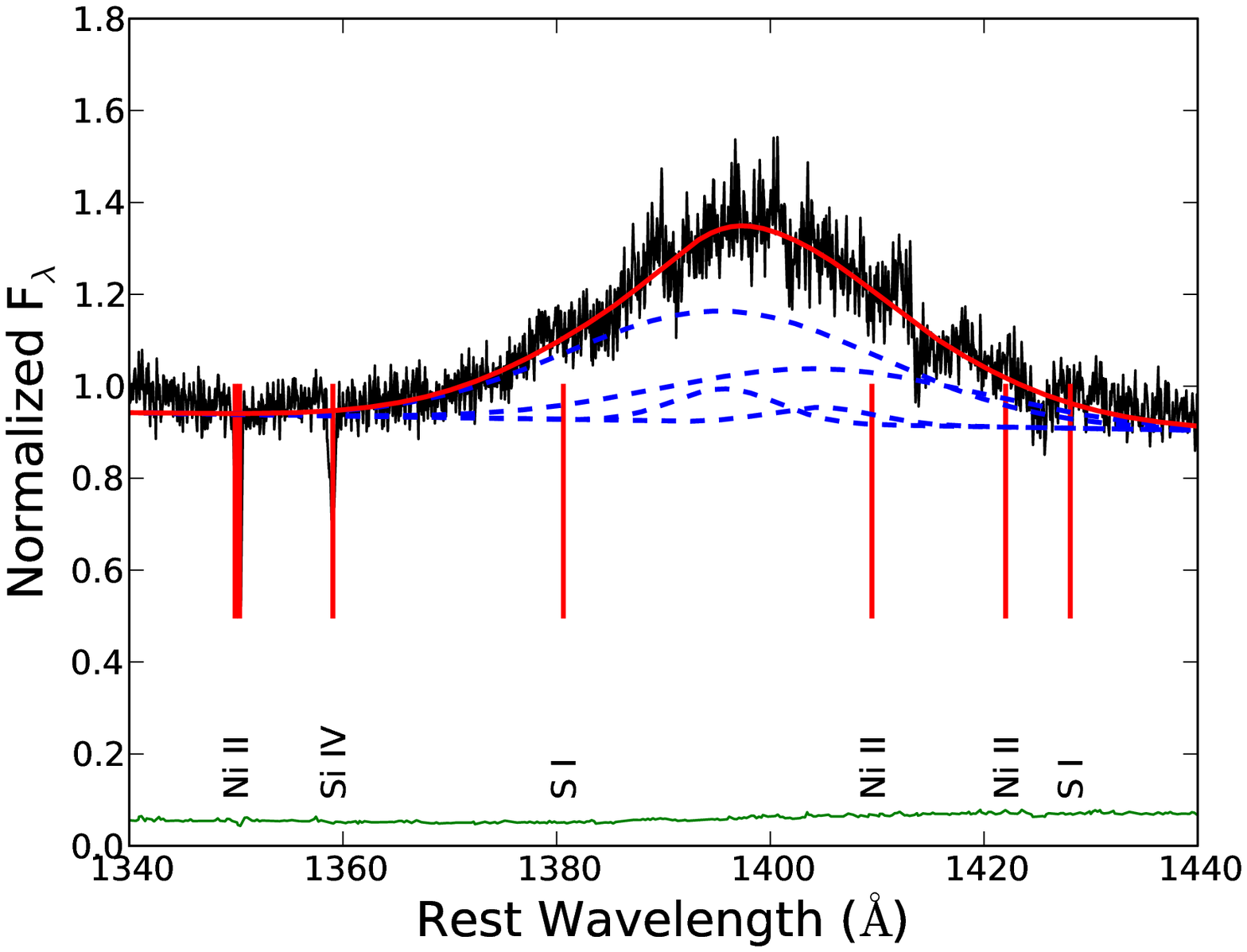} \\
\vspace{-0.07cm}
\includegraphics[width=8cm]{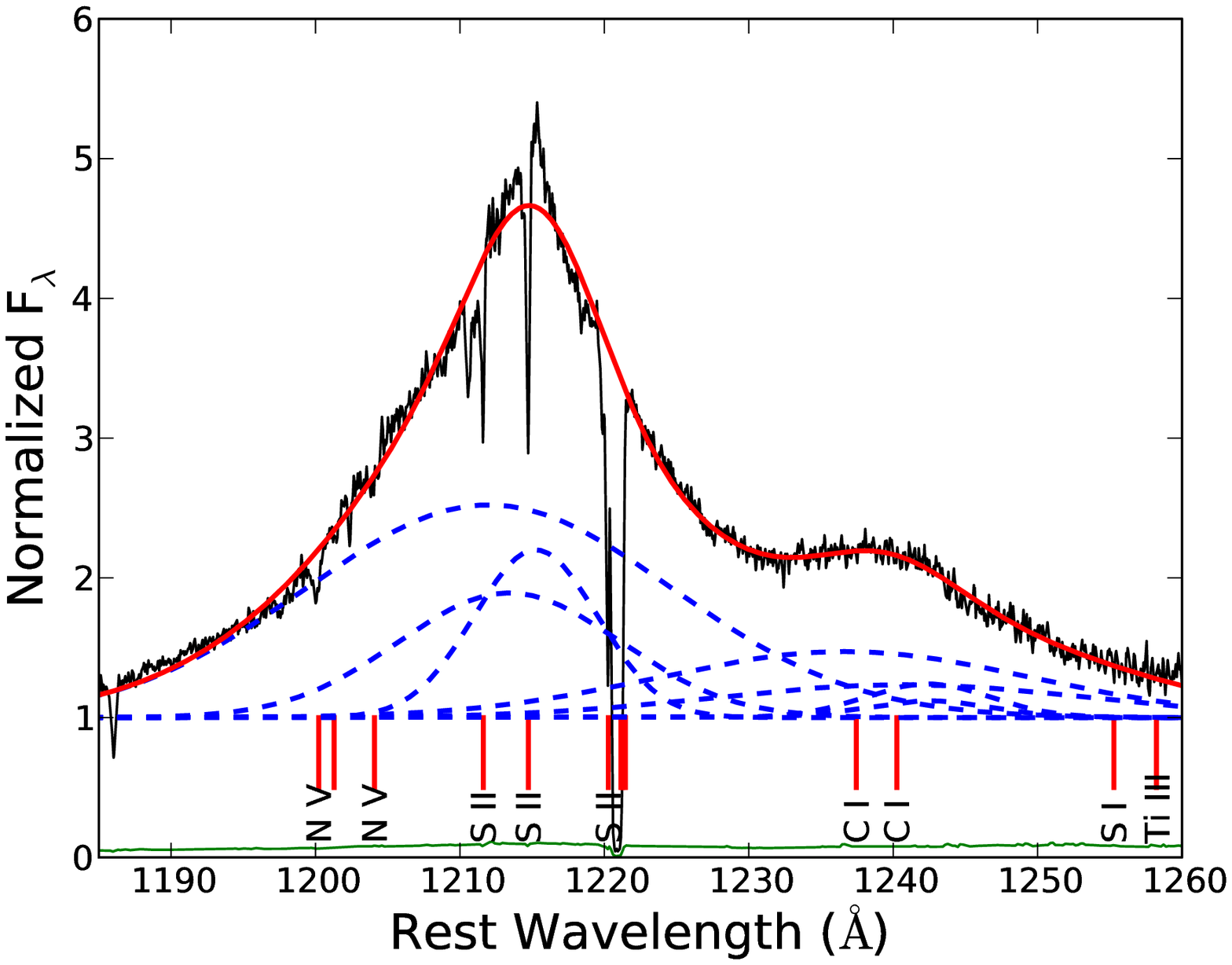}
\caption{Emission line fits to the COS ERO spectra are shown.  The \ion{C}{4} doublet is shown at top, the \ion{Si}{4} doublet, blended with \ion{O}{4}], in the middle, and Ly$\alpha$ + \ion{N}{5} in the bottom panel.   The flux (black) and errors (green), are normalized by the best-fit power law continuum.  The dashed lines show individual Gaussian components, while a solid red line shows the combined best-fit emission line model.  Galactic ISM lines are labeled; the most prominent absorption features were masked for fitting.  Additionally, the COS spectra were smoothed for illustrative purposes using moving-window averaging of four pixels. 
\label{fig-eroemission}}
\end{figure}

\begin{figure}
\centering
\includegraphics[width=8cm]{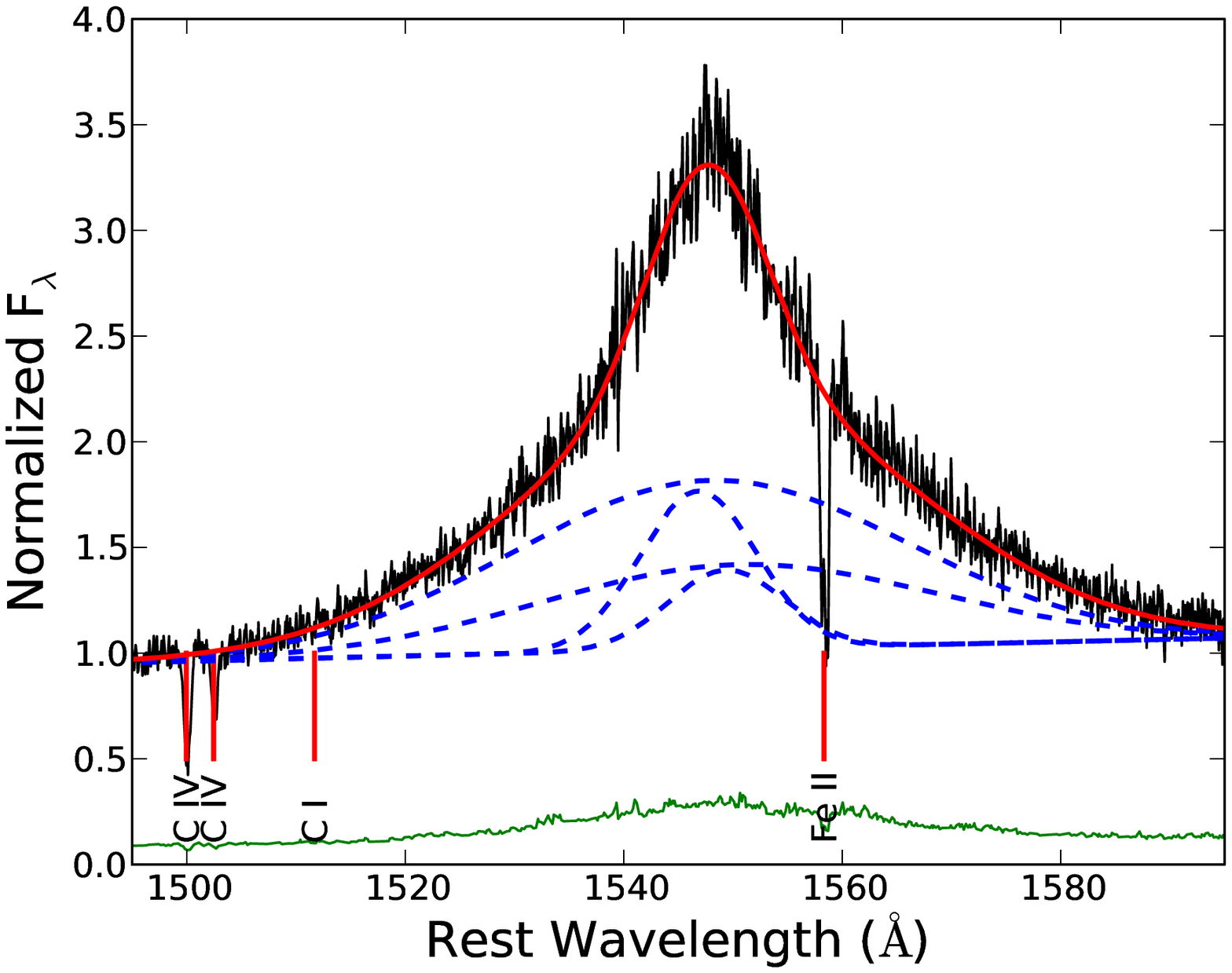} \\
\vspace{-0.07cm}
\includegraphics[width=8cm]{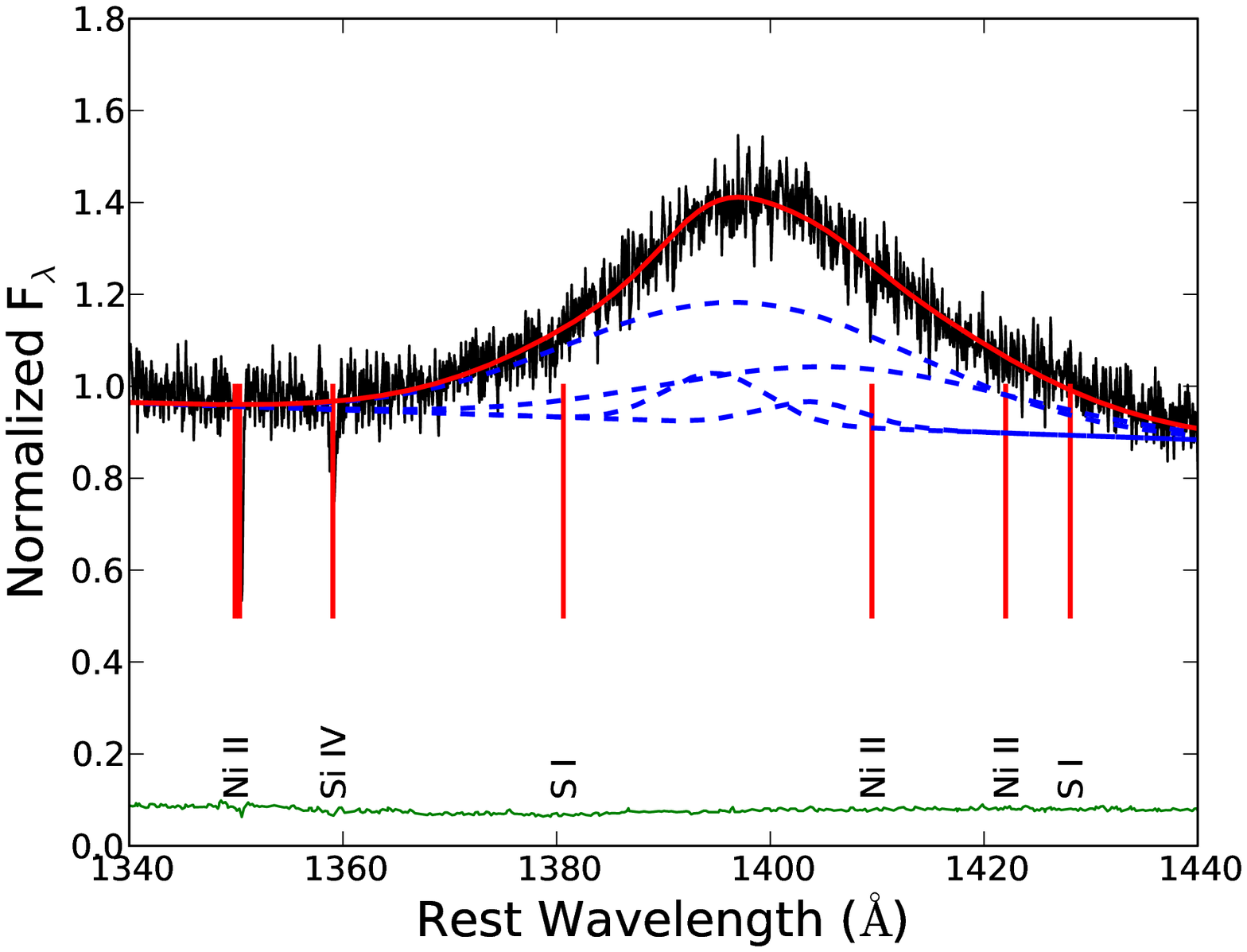} \\
\vspace{-0.07cm}
\includegraphics[width=8cm]{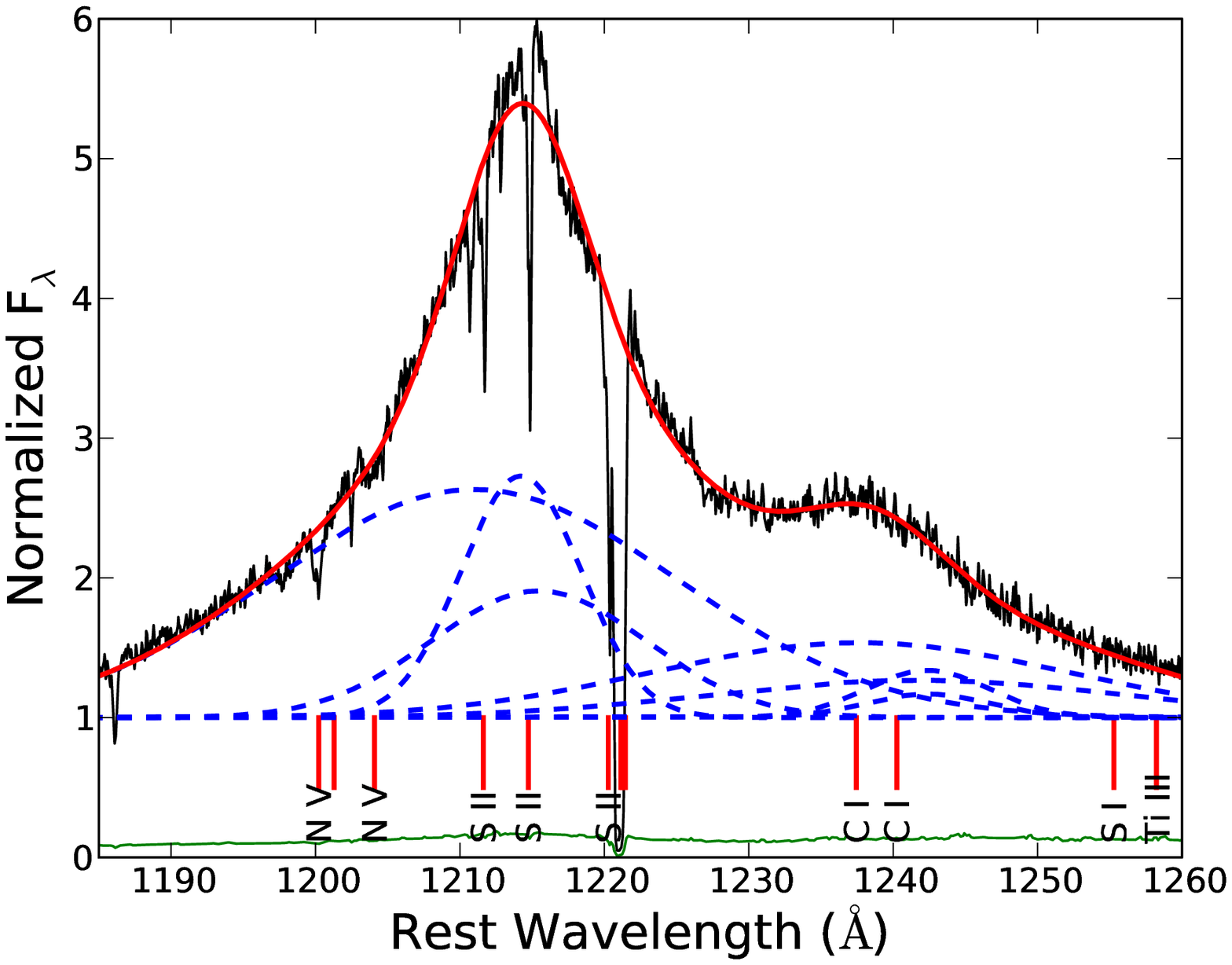}
\caption{Same as Figure~\ref{fig-eroemission}, but for the COS GTO data.
\label{fig-gtoemission}}
\end{figure}

We plot the best-fit emission line models for the ERO and GTO spectra in Figures~\ref{fig-eroemission} and \ref{fig-gtoemission}.  In general, we find that our approximations to the emission lines with Gaussian models are quite good.  The greatest discrepancies between the model and emission line-region occur, not surprisingly, in the Ly$\alpha$ profile.  For both the ERO and GTO spectra, we find that the peak of the line emission is not fully accounted for with our model.  This is due to the complications in fitting the peak emission, which is coincident with strong Galactic \ion{S}{2} absorption ($\lambda \lambda \lambda 1250.58, 1253.81, 1259.52, 1260.422$\,\AA) that had been masked from the data for the spectral fitting.

Comparing the results of our emission line fits between the ERO and GTO spectra, we find that the measured full-width-half-maximum (FWHM) values are similar for the \ion{C}{4} doublet.  We find a best-fit for both spectra with velocities of the two Gaussian components corresponding to $\sim 3000$\,km\,s$^{-1}$~and $\sim 1000$\,km\,s$^{-1}$.  
This is consistent with \ion{C}{4} FWHM values from the Palomar sample, where \citet{2005MNRAS.356.1029B} find a range of values from 1600--10250\,\km\,for their sample of 87 AGN at $z < 0.5$ AGN.

The measured velocity offsets for each of the measured line profiles do not agree, however, either between observations or different lines in the same observation.  For \ion{C}{4}, we find shifts in the line profiles from $\sim -50$\,\km\,to $\sim -340$\,\km.  These values are in line with what is found by 
\citet{2007AJ....134..294S} (who find a mean shift of $-445$ $\pm\,513$\,\km) in the Palomar-Green X-ray sample.  As found in \citet{2007AJ....134..294S}, we find that \ion{C}{4} components with larger blue-shifts also have lower $EW$ (i.e., in component 1, the GTO observation has a larger $EW$ and a smaller velocity shift than component 1 of the ERO observation).  

We find that the \ion{Si}{4} emission is red-shifted by $\sim 200-800$\,\km, while the \ion{N}{5} emission is slightly red-shifted to blue-shifted by $\sim 20$ to $-220$\,\km.  We expect error in the \ion{Si}{4} measurements, since we did not de-blend the \ion{Si}{4} and \ion{O}{4}] emission.  The \ion{N}{5} measurements suggest a smaller shift than the measured \ion{C}{4} off-sets.  

Velocity shifts in Ly$\alpha$ indicate that the strongest component (1) is blue-shifted by $\sim -1000$\,\km, with an additional red-shifted (800\,\km) component (2).  A third component, whose FWHM and EW changed slightly between observations ($\Delta FWHM \sim 100$\,\km~and $\Delta EW \sim 0.3$\AA), has a blue-shift of $-556$\km~in the ERO observation and $-57$\km~in the GTO observation.  Typical off-sets between Ly$\alpha$ and \ion{C}{4} ($v_{\rm Ly\alpha} - v_{\rm C IV}$) for radio-selected quasars are from 70--350\km \citep{1986MNRAS.218..331W}.  While we find much larger velocity shifts between Ly$\alpha$ and \ion{C}{4}, it is difficult, however, to make a direct comparison since our results are based on fits using multiple Gaussians as opposed to single Gaussian fits for each emission line.



Additionally, we find measurable differences in the emission line flux between the ERO and GTO observations.  In particular, the line fluxes are from 13--19\% higher in the ERO observation.  Thus, both the emission line flux and continuum flux are lower in the GTO observation nearly five months later.

\subsection{Comparison with GHRS and IUE Observations}\label{sect-compareuv}
In this section, we compare the COS spectra with previous observations taken in the same wavelength band.  These include a high-resolution GHRS spectrum and two low-resolution IUE spectra.  While the IUE spectra have a broader wavelength range, similar to the COS spectra, the GHRS spectrum covers only the Ly$\alpha$ region.  We include a comparison of both absorption line properties, between the GHRS and COS observations, and the emission line/continuum properties between IUE, GHRS, and COS spectra.

\subsubsection{Absorption Line Properties}~\label{sect-absorbers}
\begin{figure}
\centering
\includegraphics[width=12cm]{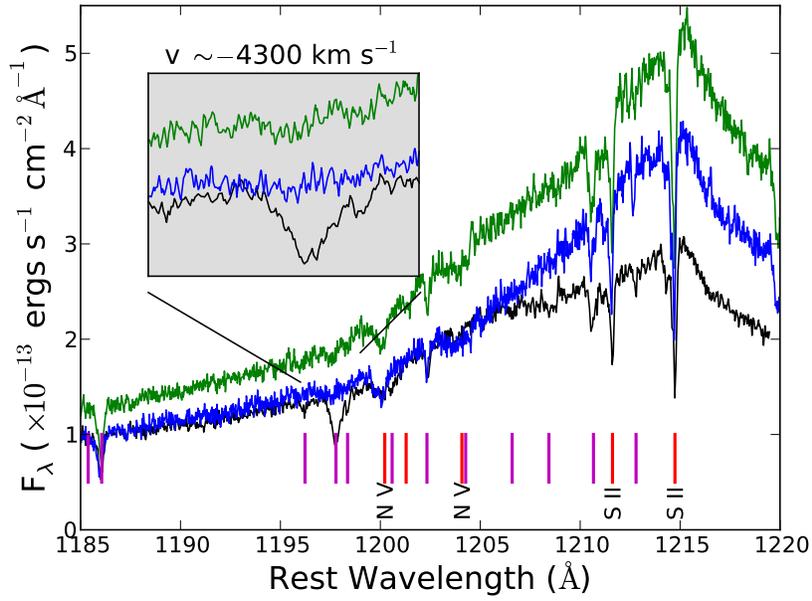}
\caption{Plotted are the Ly$\alpha$ emission line regions for the 2009 COS ERO (green) and GTO (blue) observations along with the 1997 GHRS spectrum (black).  The ERO observation had the highest flux level.  The inset shows a zoom in around the potential outflowing Ly$\alpha$ absorption feature that was detected in the GHRS spectrum, as well as the four 2000-2001 FUSE observations.  Clearly, the feature is weaker in the COS observations (by a factor of $\sim 5$).  Additional absorption features in the spectrum include Galactic ISM features (red) and HVCs/additional weak intrinsic absorbers (magenta), which are detailed in \citet{2000ApJS..130..121P},  \citet{2003ApJ...585..336C}, and this paper.  See Figure~\ref{fig-intrinsic} and the text for more details on these features.
\label{fig-comparelya}}
\end{figure}

To compare the COS observations with the GHRS spectrum, we first applied the Galactic reddening correction using the \citet{1989ApJ...345..245C} reddening curve and shifted the spectrum to the AGN rest frame.  In Figure~\ref{fig-comparelya}, we plot the Ly$\alpha$ emission line region for the COS and GHRS observations.  Several weak absorption components, in addition to Galactic absorption (e.g., from \ion{S}{2}, \ion{Si}{2}, and \ion{N}{5}), are detected in both the COS and GHRS observations.

The major difference between the COS and GHRS observations of Ly$\alpha$ is that an intrinsic absorber (interpreted as a signature of an outflow emanating from the AGN, e.g.,
\citealt{1997AJ....113..136B}, section 1), 
 clearly visible in the GHRS data, is present only at a very low level in the COS observations (see Figure~\ref{fig-comparelya}).  To further quantify the properties of this absorption feature, we include the parameters from a Voigt profile fit to the GHRS, individual COS observations, and combined COS observations in Table~\ref{tbl-absorber}.  We find that the velocity of this feature ($\sim -4300$\km) is roughly consistent with the absorber found in the 2000-2001 FUSE observations ($\sim -4100$\km).  There is no significant difference in the measured FWHM of this feature between the GHRS and COS observations, but a slight ($\approx -10$\km) increase in the velocity offset.  However, the equivalent width of the line decreased by a factor of $\sim 5$ (from $EW \sim 200$\,m\AA\ in 1997 to $EW \sim 35$\,m\AA\ in 2009) between the GHRS and COS observations.  We find no difference in this feature between the COS ERO and GTO spectra and therefore include in the table measurements from a combined COS ERO + GTO fit to this feature (see below for our procedure).

\begin{figure} 
\epsscale{1}
\plotone{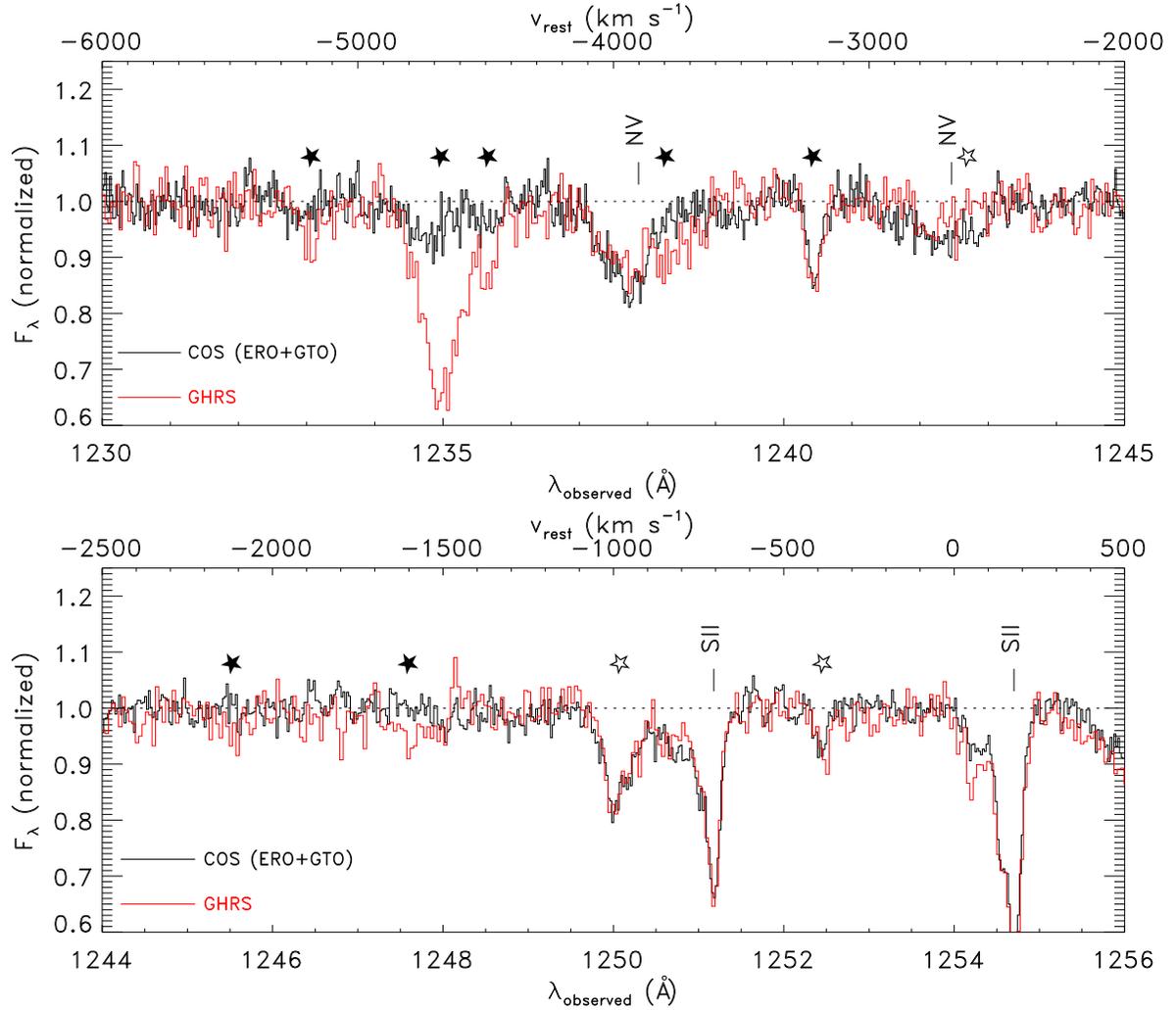} 
\caption{Comparison of the normalized GHRS (red) and COS ERO$+$GTO data (black) is shown.  Ly$\alpha$ absorption features noted by Penton et al. (2000) are marked with filled stars and open stars mark three additional features noted here (two absorbers, $\lambda_{\rm obs}=1223.5$\,\AA\,and $1224.2$\,\AA, are not shown).  Absorption arising from interstellar N\,V and S\,II lines are marked.  Several features appear unchanged between observation epochs while several others exhibit marked changes.  See text for discussion of individual systems.
\label{fig-intrinsic}}
\end{figure}

Additionally, we searched for other absorbers in the GHRS and COS datasets.  Since many of these lines are quite weak, we maximize our data quality by merging the COS ERO and GTO data into a single combined spectrum; the flux in the GTO data was multiplicatively scaled by 1.4 to match the mean continuum level seen in the ERO data.  The flux in both spectra were normalized using low-order Legendre polynomials fitted to short regions of the spectrum ($\sim10-20$ \AA).  Figure~\ref{fig-intrinsic} compares the absorption lines in the GHRS and combined COS data.  
\citet{2000ApJS..130..121P} identified nine absorption lines, including the absorber described above, (filled stars) in the GHRS data at $1222$\,\AA\,$<\lambda_{\rm obs}<1247.3$\,\AA, corresponding to Ly$\alpha$ absorbers at $|v|>1500$ \km.  We note three additional absorbers that were not identified in \citet{2000ApJS..130..121P} (open stars), including one feature which appears in the COS data only.  



We present measurements of the twelve detected Ly$\alpha$ absorbers in Table~\ref{tbl-intrinsic}, including measurements of the $\sim -4300$\km ($cz_{AGN} = -4250$\km) listed in Table~\ref{tbl-absorber} for a clear comparison with the additional weak absorbers.  An adjacent, weaker absorption component to the $-4300$\km\,absorber is found at $\lambda_{\rm obs}=1236.9$\,\AA\,($v=-4100$ \km).   It is likely associated with the $-4300$\km\,feature, as its $EW$ appears similarly diminished in the COS observation.  An additional weak absorber, coincident with a second `outflow' component in the {\it FUSE} observations, also appears to be present in the GHRS and COS spectra at $\lambda_{\rm obs} = 1239.2$\AA.  In \citet{2008AJ....136.1201D}, this second component is shown to arise only in the last of the four {\it FUSE} observations, with an equivalent width of 0.12\AA~and velocity of $-3680$\km.  The corresponding component in our spectra is blended with Galactic \ion{N}{5} absorption, but present, with a velocity of $\sim -3550$\km.  Additionally, the strength of this feature appears to have decreased between the GHRS and COS observations.  We estimate an equivalent width for the blended absorption line of $54 \pm 8$m\AA~in the 1997 GHRS spectrum and $14\pm 5$m\AA~in the 2009 COS spectra.  Since this feature is blended with \ion{N}{5} Galactic absorption, the measurements are uncertain, but it appears that the decrease from 1997 to 2009 is consistent with the factor of 5 decrease in the absorbers at $-4250$\km and $-4100$\km.

Of the additional Ly$\alpha$ absorbers detected, four absorbers are consistent between GHRS and COS ($\lambda_{\rm obs} = 1224.2$\AA, 1241.0\AA, 1249.6\AA, and 1251.8\AA).  These absorbers are likely interstellar features from high-velocity clouds (as described in \citealt{2000ApJS..130..121P}).  An additional four absorbers are detected but at a low-significance ($< 10\sigma$).  Finally, the feature at $\lambda_{\rm obs} = 1243.0$, also blended with a Galactic \ion{N}{5} absorption line, does not appear in the GHRS data but is seen in the combined COS spectrum.  Thus, this is a fourth weak intrinsic absorber in the AGN spectrum.


If the weak intrinsic absorption features seen in the {\it FUSE}, GHRS, and COS spectra are associated with an ionized outflow, we expect a correlation between changes in the strengths of the lines and the luminosity of the source.  Particularly, for a factor of five change in the $-4300$\km~absorber, we expect a significant change in the continuum or line flux between the GHRS and COS observations.  However, this is clearly not the case, as seen in Figure~\ref{fig-comparelya}.  Either a change in the covering fraction or column density of a cloud in the line of sight is a more likely cause of the changes in the intrinsic absorbers.  This could be an effect of bulk motions of a cloud/clouds in the AGN host galaxy.


\subsubsection{Continuum and Emission-Line Properties}\label{sect-ghrsiue}

In order to obtain an estimate of the continuum and Ly$\alpha$ line flux in the GHRS observation, we fit the Galactic reddening-corrected spectrum with a combination of three Gaussians and a power law continuum ($a_0 \times (\lambda/1750$\AA$)^{-\alpha}$).  As for the COS spectra, we masked prominent features, including the intrinsic absorber.  In order to allow some consistency between the COS and GHRS spectra, we fixed the power-law index to the average value from the COS spectra ($\alpha = 1.30$).  We also constrained the velocity offset and width of the first two Gaussians to the values obtained in the COS ERO observation, allowing the amplitude and the parameters of a third Gaussian to vary.  Our best fit continuum flux density at 1750\AA\,was $a_0 = (0.511 \pm 0.002) \times 10^{-13}$\,\flux\,\AA$^{-1}$.  The flux of the first Gaussian component is $6.76 \times 10^{-12}$\,ergs\,s$^{-1}$\,cm$^{-2}$, while the second Gaussian component had a flux of zero. The third Gaussian component had a velocity offset of $-276.2 \pm 5.9$\km, FWHM of $362.6 \pm 7.1$\km, and line flux of $2.57 \times 10^{-13}$\flux.  The complex shape of the line, essentially modeled by two Gaussians, was not fit well with this model ($\chi^2/{\rm dof} = 3.6$), but this simple characterization is sufficient to compare the flux from the GHRS spectrum to that in the COS spectra.  The total measured line flux is $7.0 \times 10^{-12}$\,\flux.  

To compare the GHRS results with the COS observations, we must determine the parameters for the continuum and Ly$\alpha$ emission using the same narrow band used to fit the GHRS.  When we re-compute the COS continuum and Ly$\alpha$ properties using the same model applied to the GHRS, we find $a_0 = (0.554 \pm 0.002) \times 10^{-13}$\,\flux\,\AA$^{-1}$~and a total Ly$\alpha$ line flux of $8.8 \times 10^{-12}$\,\flux~for the ERO observation and $a_0 = (0.468 \pm 0.003) \times 10^{-13}$\,\flux\,\AA$^{-1}$~and a total Ly$\alpha$ line flux of $6.4 \times 10^{-12}$\,\flux~for the GTO observation.  These results confirm what a visual inspection of Figure~\ref{fig-comparelya} suggests: the ERO observation is the brightest observation in both continuum flux and Ly$\alpha$ emission.  The GHRS measured continuum normalization is 8\% lower than the ERO value and 8\% higher than the GTO value.  The Ly$\alpha$ emission from the GHRS observation is 20\% lower than in the ERO observation and only about 9\% higher than in the GTO observation.  This shows that there is no drastic change in flux between the COS and GHRS observations and confirms that the drastic change in the equivalent width of the $-4250$\km~absorber is probably not directly related to the relative luminosity of the source, either in the continuum or the Ly$\alpha$ emission line.

\begin{figure}
\centering
\includegraphics[width=10cm]{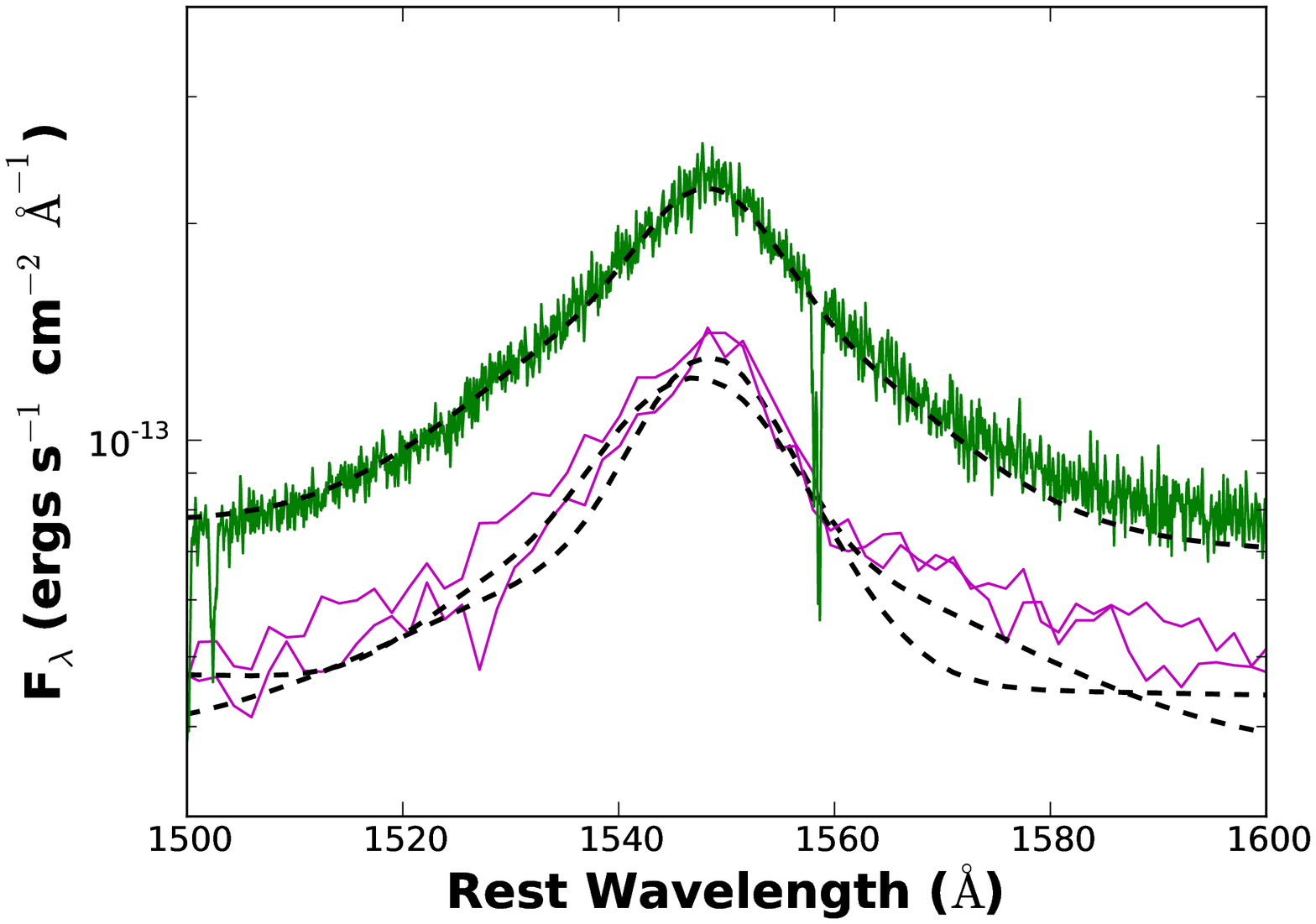}
\includegraphics[width=10cm]{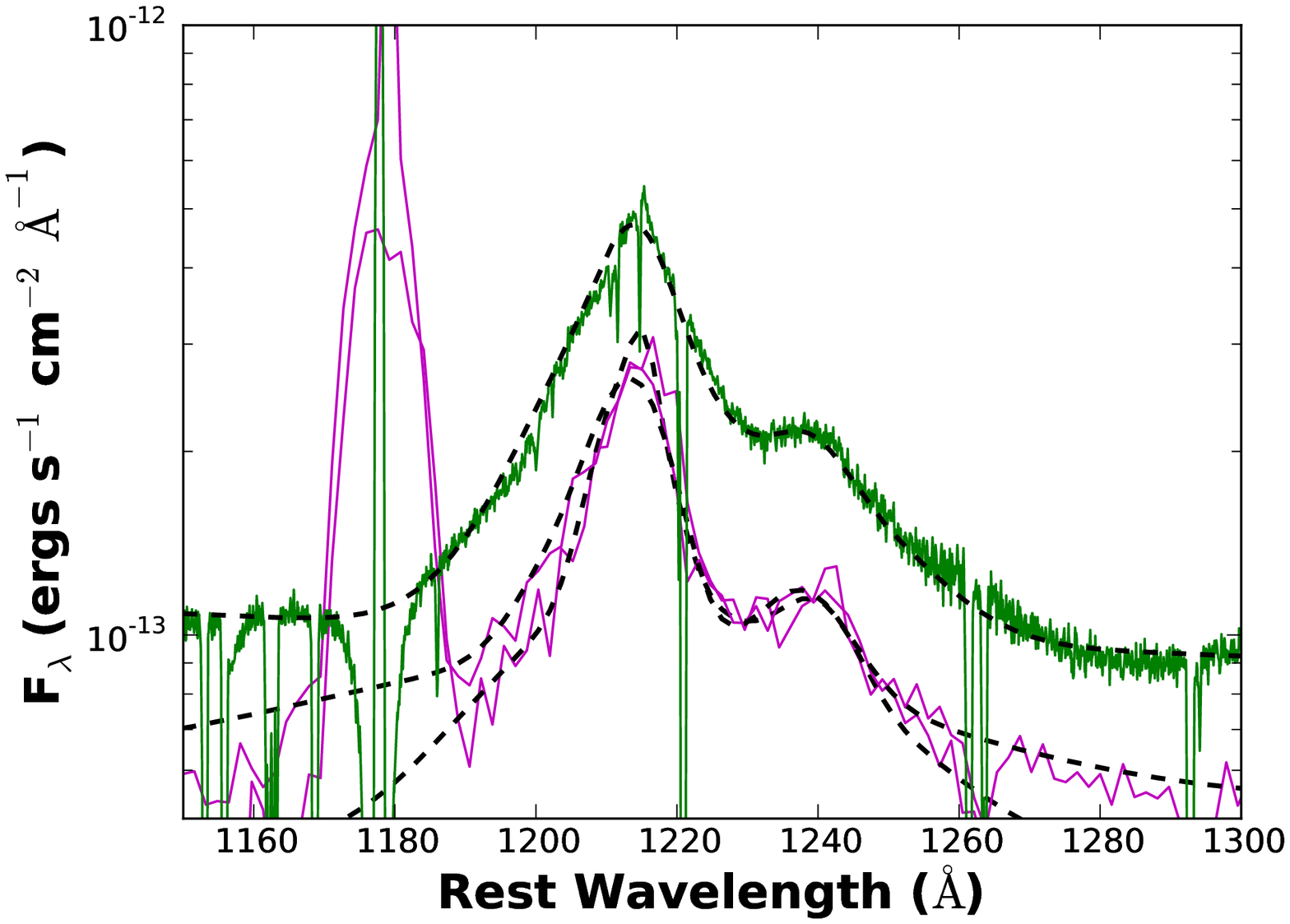}
\caption{Plotted is a comparison of the spectra and best-fit model for the COS ERO (green) and IUE (magenta) spectra in \ion{C}{4} and the Ly$\alpha$ + \ion{N}{5} regions.  The best-fit emission line fits are shown in black.  The COS spectra are smoothed by moving-window averaging of four pixels.  Low-resolution in the IUE spectrum coupled with strong geo-coronal Ly$\alpha$ emission make it difficult to compare the line flux in Ly$\alpha$ + \ion{N}{5} doublet with the higher resolution COS spectra.
\label{fig-IUE}}
\end{figure}

Next, we describe our analysis of the IUE spectra and compare their properties to the COS and GHRS spectra.  The resolution of the IUE spectra ($R \sim 200$) of Mrk 817 is unfortunately too low to search for outflow signatures.  It is, however, sufficient to obtain the continuum and emission line properties of prominent lines.  To determine these parameters, we first corrected the two IUE spectra for the effects of Galactic reddening with the \citet{1989ApJ...345..245C} reddening curve.  We measured the continuum parameters in the same way as for the original fits to the COS spectra, choosing continuum regions free of absorption/emission features with which to fit a power law model.  We also followed the same procedure to fit the emission lines as for the COS spectra -- fitting the \ion{C}{4} emission with a broad and intermediate width line and applying the same FWHM measurements to the \ion{Si}{4} (+ \ion{O}{4}]), \ion{N}{5}, and Ly$\alpha$ (two of three Gaussians) emission.  The measured parameters (power-law parameters and total line flux) are shown in Table~\ref{tbl-iuespectra}.  In Figure~\ref{fig-IUE}, we plot the IUE spectra and their best-fit emission line fits for the \ion{C}{4} doublet and Ly$\alpha$ + \ion{N}{5} doublet along with the COS ERO spectrum.

\begin{figure}
\centering
\includegraphics[width=12cm]{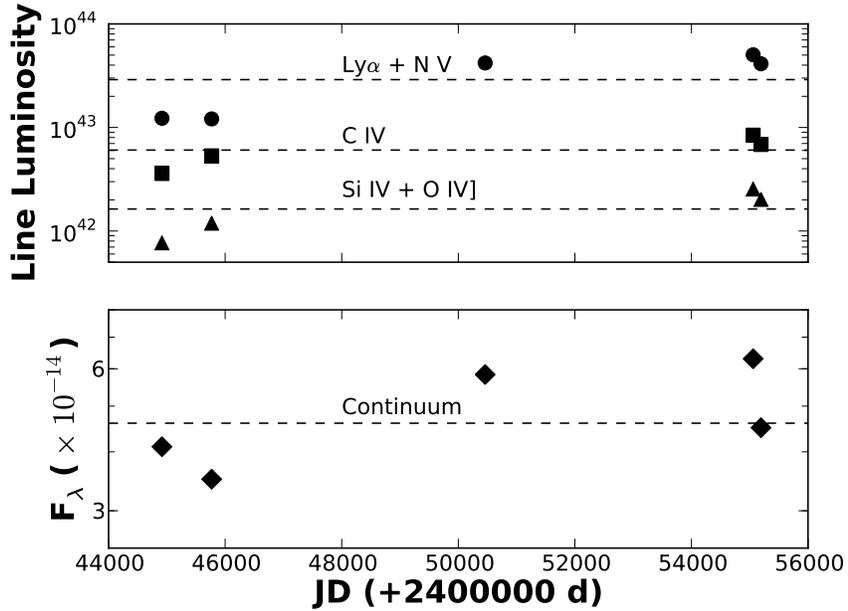}
\caption{The UV light curve for Mrk 817 is plotted, showing changes in luminosity (in ergs\,s$^{-1}$) for the emission lines (Ly$\alpha$ + \ion{N}{5} doublet, \ion{C}{4} doublet, and the \ion{Si}{4} doublet, blended with \ion{O}{4}]) in the top panel and the continuum flux density (the normalization value $a_0$ from the power law fits) in the bottom panel.  The horizontal lines show the averages for the IUE and COS observations in each of the values.  The GHRS measurements are estimates based on spectral fitting in the Ly$\alpha$ range, scaling to the broader bandpass fits of the COS ERO observations as explained in the text.  We find the IUE observations to show lower continuum and line luminosities than the COS and GHRS observations.  However, there are no drastic changes over the observations, which span almost 30 years.
\label{fig-uvlc}}
\end{figure}

In order to compare our results from the IUE, GHRS, and COS spectra, we plot a light curve of the  line luminosities and continuum flux densities in Figure~\ref{fig-uvlc}.  Since the GHRS spectra only cover Ly$\alpha$, we needed to make a few assumptions to include the GHRS flux for the continuum and Ly$\alpha$ + \ion{N}{5}.  We assumed that the ratio between the GHRS and COS ERO measurements of the continuum and Ly$\alpha$ emission from fits of the narrow region from 1185.0--1219.5\,\AA~ were the same as would be measured for the broad-band continuum and Ly$\alpha$ + \ion{N}{5}.  Therefore, the GHRS points correspond to the COS ERO values from Tables~\ref{tbl-powerlaw} and \ref{tbl-emission-ero} divided by the factor of 1.08 (continuum) and 1.20 (Ly$\alpha$) found from our narrow-band fits.  The IUE values are from Table~\ref{tbl-iuespectra}.  We combine Ly$\alpha$ and \ion{N}{5} to minimize errors associated with separating the blended lines.

Our results show that the emission line luminosities and continuum flux density levels of the IUE observations are lower than those in the GHRS and COS observations.  While the Ly$\alpha$ + \ion{N}{5} doublet appears to be at a much lower level in the IUE spectra, the lower measurement is more likely due to the larger uncertainty associated with the lower-resolution data coupled with the complexity of the emission lines and the presence of a strong geo-coronal Ly$\alpha$ feature (see Figure~\ref{fig-IUE}).  Comparing the more reliable measurements of the continuum and \ion{C}{4} doublet, we estimate a change of approximately 1.5--1.8 and 1.6--2.3, respectively.  Therefore, there appears to be no drastic (i.e., order of magnitude) change in the UV continuum or emission lines of Mrk 817 over the observations -- spanning nearly 30 years.   

\section{The X-ray Spectra}~\label{sect-xray}
In this section, we describe the results of our analysis of the observations of Mrk 817 from XMM-Newton and Swift.  We include the observation details and reduction of the X-ray spectra in \S~\ref{sect-xrayobservations}. 
In \S~\ref{sect-xmmspectra}, we include spectral fitting of the XMM-Newton EPIC spectra, with the specific goal of determining whether an ionized absorber is present.  In \S~\ref{sect-xrayvariability}, we search for variability in the spectra, as well as the associated optical/UV photometry, between the XMM-Newton and Swift observations.

\subsection{Observations}\label{sect-xrayobservations}
Serendipitously, the X-ray observatory XMM-Newton \citep{2001A&A...365L...1J} observed Mrk 817 within 2 weeks of the HST COS GTO observation (the second of the two COS observations).  Mrk 817 was observed on 2009 December 13 for 14220\,s, under the program of PI Brandt (observation ID: 0601781401).
The data were reduced according to the standard guidelines in the XMM-Newton User's Manual\footnote{{\tt \url{http://heasarc.nasa.gov/docs/xmm/sas/USG/}}}, using
the XMM-Newton Science Analysis Software ({\sc{sas}}) version 9.0.0.  We analyzed data from the two types of European Photon Imaging Camera (EPIC) onboard XMM-Newton: the two front-illuminated EPIC Metal Oxide Semi-conductor (MOS; \citealt{2001A&A...365L..27T}) CCD arrays and the back-illuminated EPIC pn \citep{2001A&A...365L..18S} CCD array.  The EPIC cameras are sensitive in the $\sim 0.3-10$\,keV band and have a spectral resolution of $E/\Delta E \sim 20$--$50$ with an angular resolution of $\sim 6$\arcsec\,$FWHM$.

The tasks {\sc{epchain}} and {\sc{emchain}} were used
to reduce the data from the pn and MOS instruments, respectively.  Initially a circular source region of
radius 36\arcsec\,was used to extract a source spectrum; however, the {\sc{sas}} {\sc{epatplot}} task revealed
that the observation exhibited ``pile-up'' in the pn and MOS1 detectors.  Pile-up occurs when two or more X-ray photons deposit charge in a single pixel or neighboring pixel during one read-out cycle and are therefore registered as a single event with an energy equal to the sum of the photon energies.  To correct for pile-up, we followed the recommended approach of using an annular
source region (inner radius 15\arcsec, outer radius 45\arcsec\,for the pn; the corresponding radii for MOS1 were 7.5\arcsec\,and 45\arcsec; and the MOS2 data did not display pile-up) to extract the spectra for those detectors, until {\sc{epatplot}} displayed no strong signature of
pile-up\footnote{See the following thread on dealing with pile-up in the EPIC detectors: \url{http://xmm.esac.esa.int/sas/current/documentation/threads/epatplot.shtml}}.  Background regions were either chosen to be circles near the source or annuli that exclude the central source. Additionally, the background light curves (between 10--12 keV) were inspected for flaring, and a comparison
of source and background light curves in the same energy range was used to determine the portions of the observation
in which the background was sufficiently low compared to the source.  Approximately 75\% of the pn observation and 93\% of the MOS observations were deemed free of flaring; the subsequent spectra were generated from the usable portions of each observation.  Response matrices and auxiliary files were extracted using the tools {\sc{rmfgen}} and {\sc{arfgen}}, and the final spectra were grouped with a minimum of 20 counts per bin using the {\sc{grppha}} tool.  

The pn spectrum has an exposure time of 7257\,s with an average count rate of 4.45\,cts\,s$^{-1}$.  The MOS spectra have exposure times of 5882\,s (MOS1) and 5858\,s (MOS2), with average count rates of 1.95\,cts\,s$^{-1}$ (MOS1) and 3.02\,cts\,s$^{-1}$ (MOS2).  The difference in the MOS1 and MOS2 count rates is due to the different apertures required since pile-up is detected in MOS1 but not MOS2.  No significant rapid variability is seen in the pn or MOS light curves.

In addition to the XMM-Newton spectrum,  we include 5 lower signal-to-noise ratio X-ray spectra observed with the Swift X-ray telescope (XRT) \citep{2000SPIE.4140...64B,2000SPIE.4140...87H}.  The Swift XRT is sensitive in a similar energy band to the XMM-Newton pn and MOS CCDs ($\sim 0.3$--10\,keV).  The XRT observations were all downloaded from the High Energy Astrophysics Science Archive Research Center\footnote{\url{http://heasarc.gsfc.nasa.gov/}}.  We use the processed clean event files from the photon counting mode to create the spectra.  The source spectra were extracted from a circular aperture with a 75\arcsec\,radius, centered on the source, using the FTOOL {\tt XSELECT}.  The background spectra were extracted from a 100\arcsec\,circular region near the position of Mrk 817 and free of additional sources.  Standard response and ancillary response files from the Swift XRT calibration database were used for the photon counting event files with grades 0-12.  Details of the XRT observations are included in Table~\ref{tbl-xrt}.

\subsection{XMM-Newton Spectral Analysis}\label{sect-xmmspectra}
Although Mrk 817 is a bright X-ray source, the only broad-band (0.3--10\,keV) X-ray spectra taken to date are a series of short ($\le 6$\,ks) snapshots with the Swift XRT (which we discuss in the following section) and the 14\,ks XMM-Newton observation we now present.  The XMM-Newton observation of Mrk 817 was obtained as part of a program to study the X-ray properties of a uniform sample of AGN detected in the 22-month Swift BAT catalog \citep{2010ApJS..186..378T}.  In this paper, we present a basic analysis of the EPIC spectra -- principally to determine whether an X-ray outflow is detected.  A more detailed analysis will be presented in Vasudevan et al. (in prep).  

Using XSPEC 12.5  \citep{ADASS_96_A}, we simultaneously fit the pn, MOS1, and MOS2 X-ray spectra with a simple absorbed power law.  We accounted for Galactic absorption with the {\tt tbabs} model, using the \citet{1990ARAA..28..215D} value of N$_{\rm H} = 1.15 \times 10^{20}$\,cm$^{-2}$.  We found that the addition of intrinsic neutral absorption from the AGN did not improve the fit.  The residuals to this model did, however, show a soft excess, which we fit with a black body model.  This simple blackbody + power-law (where the power-law model corresponds to $F(E) \propto E^{-(\Gamma - 1})$) model is statistically a good fit to the data, with reduced $\chi^2/dof = 1.03/399$.  The best-fit parameters of this model include $\Gamma = 2.17 \pm 0.06$, with a normalization of $6.14 \pm 0.19 \times 10^{-3}$\,photons\,cm$^{-2}$\,s$^{-1}$\,keV$^{-1}$ at 1\,keV, and  $kT = 0.099^{+0.003}_{-0.011}$\,keV, with a normalization of $8.4^{+1.7}_{-2.1} \times 10^{-5}$\,photons\,cm$^{-2}$\,s$^{-1}$\,keV$^{-1}$ at 1\,keV. Quoted errors here correspond to the 90\% confidence level. The corresponding absorption corrected flux in the 0.4--10\,keV band is $3.4 \times 10^{-11}$\,\flux.

Sources that exhibit X-ray signatures of outflows are also known to show UV absorbers 
(e.g., \citealt{1999ApJ...516..750C}).  A simple picture of gas in a single ionization state producing both the UV and X-ray absorber features is disproved by simultaneous UV/X-ray grating observations.  However, these observations have revealed that the velocity field of the UV and X-ray absorbers are similar for a few sources (for instance, see \citealt{2002ApJ...574..643K} for details on NGC 3783).  Thus, the UV absorbers are thought perhaps to arise in higher density knots within the X-ray outflow \citep{1995ApJ...447..512K}.  Since no UV absorption signatures were seen in the ionized lines (\ion{N}{5} and \ion{C}{4}) in the COS GTO spectrum, we expect to find no outflow signatures in the X-ray spectrum of Mrk 817. 

\begin{figure}
\centering
\includegraphics[width=14cm]{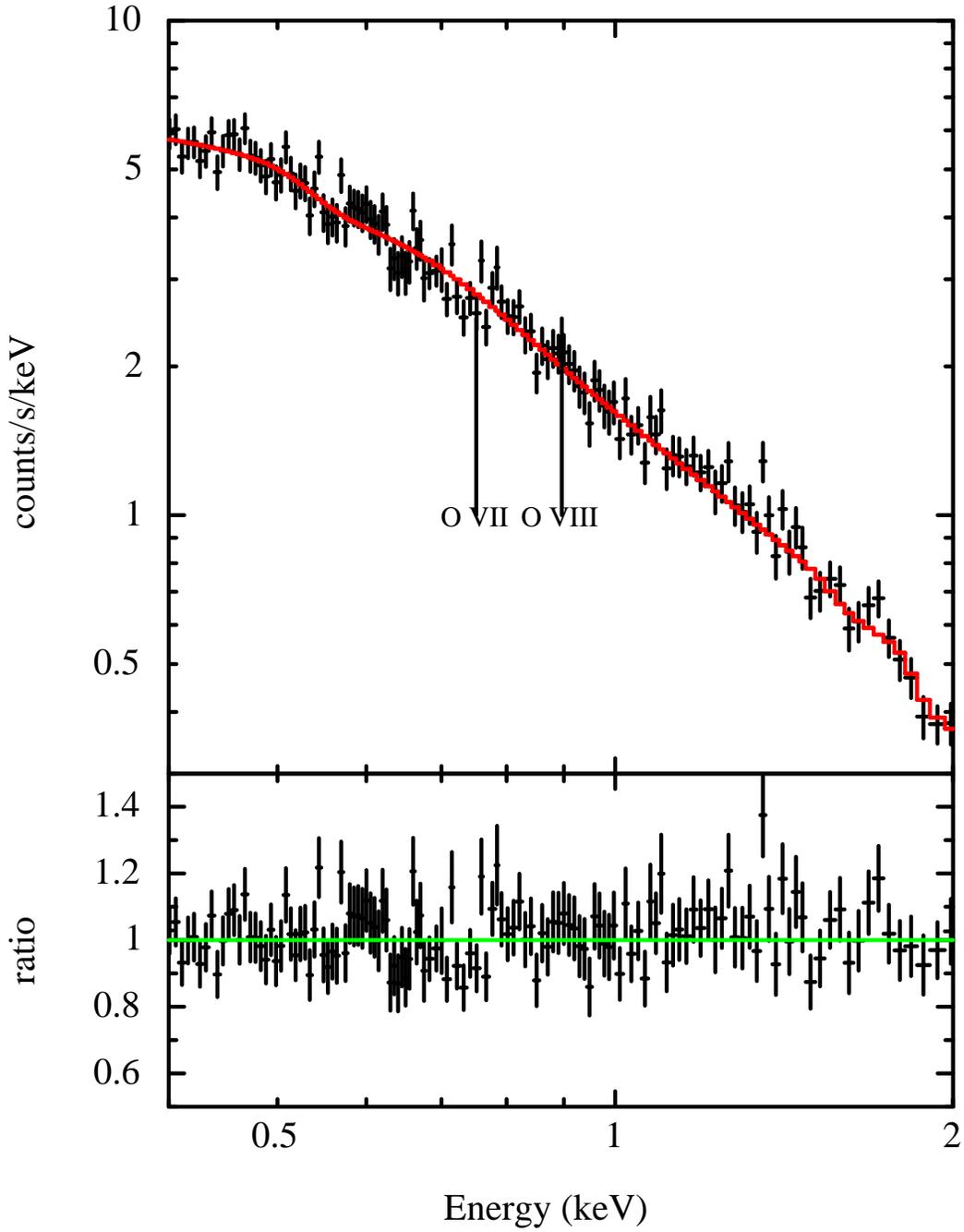}
\caption{The XMM-Newton pn spectrum of Mrk 817 is shown in the observed 0.4--2\,keV band, re-binned to a signal-to-noise ratio of 10.  This X-ray observation was taken within two weeks of the COS GTO observation.  The bottom panel shows the ratio of the data/model, where the model is a blackbody + power-law model.  The low-energy spectrum exhibits no signs of an outflow, which is what we expected based on the lack of a strong detection in the UV COS observations.   The location of the \ion{O}{7} and \ion{O}{8} absorption edges are indicated on the plot.
\label{fig-xray}}
\end{figure}

 X-ray outflow signatures are typically detected in CCD data through the \ion{O}{7} and \ion{O}{8} absorption edges at energies of 0.73\,keV and 0.87\,keV, respectively.  As shown in Figure~\ref{fig-xray}, which presents the pn spectrum in the soft X-ray band and the ratio of data/model in the same band, there is no evidence for the presence of an X-ray warm absorber/outflow.  To test this result statistically, we added two {\tt zedge} models, one to account for each of the absorption edges, at the energies of the edges.  The change in $\chi^2$ upon adding the edges is very low ($\Delta\chi^2 = 0.58$).  Calculating errors on the optical depth from this model, we find upper limits of $\tau_{\rm O VII} < 0.11$ and $\tau_{\rm O VIII} < 0.05$ at the 90\% confidence level, confirming that outflows are not detected in the XMM-Newton X-ray spectrum of Mrk 817.  



\subsection{Variability}\label{sect-xrayvariability}

\subsubsection{Long-term variability from the X-ray observations}
While the sensitivity of the Swift XRT observations is too low to search for outflows, the XRT spectra are useful for determining changes in both the X-ray flux and overall spectral shape over the 2.5\,yr period covered with the Swift XRT and XMM-Newton observations.  To search for variability between the XMM-Newton and Swift XRT observations, we followed a similar procedure to \citet{2008ApJ...674..686W}, who characterized X-ray variability  in 22 Seyferts detected in the Swift Burst Alert Telescope AGN survey \citep{2008ApJ...681..113T} through simultaneous fits to Swift XRT and XMM-Newton spectra.  We use the best-fit blackbody + power-law model from the XMM-Newton spectra as a baseline model to search for variability between the pn and the five Swift XRT spectra.

\begin{figure}
\centering
\includegraphics[width=12cm,angle=270]{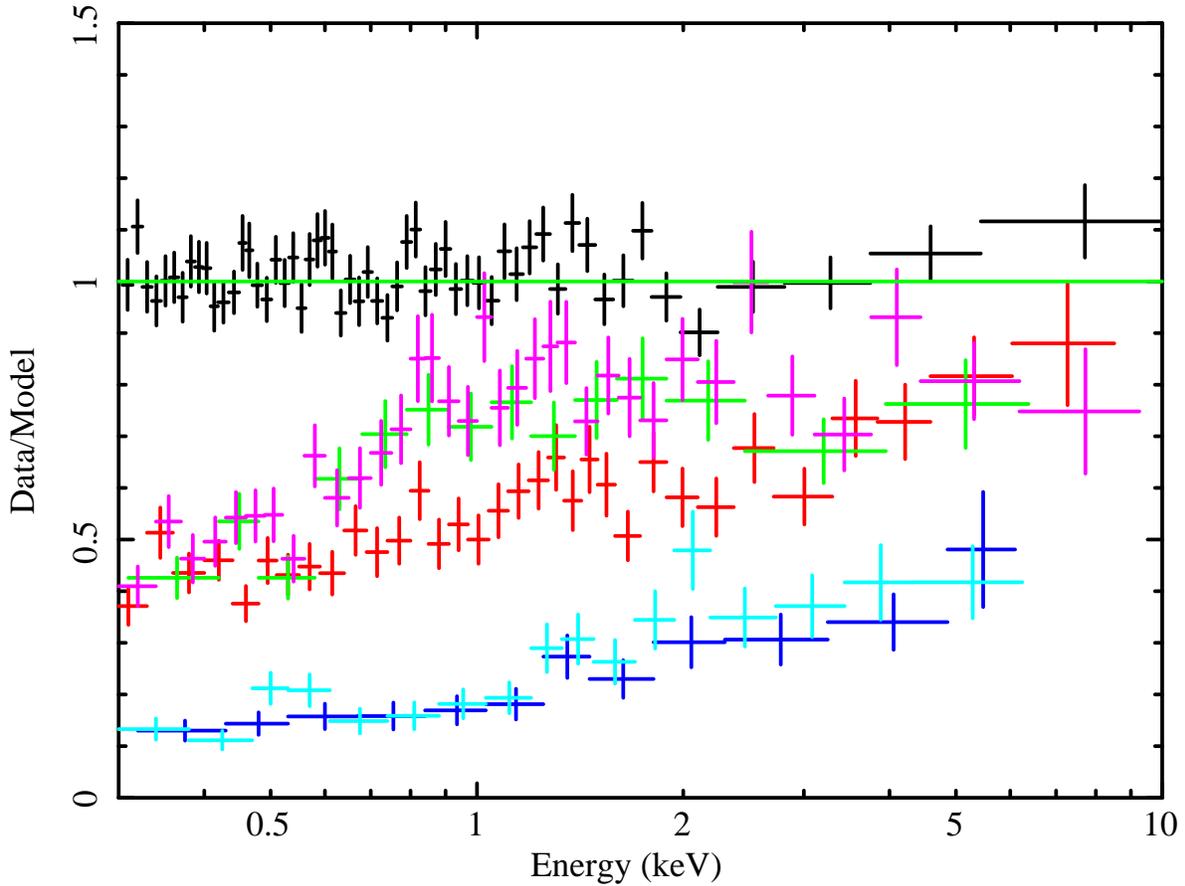}
\caption{Plotted is the ratio of the data/model for the blackbody + power-law model fit to the December 2009 XMM-Newton pn spectrum (black, re-binned to a signal-to-noise of 20 for illustrative purposes) and the five Swift XRT spectra (re-binned to a signal-to-noise from 5 to 10, depending on the exposure time of the observation).  The model applied to the spectra is the best-fit model to the pn spectrum.  It is clear that the source varied both in flux and spectral shape between the X-ray observations.  The lowest flux points (both in blue) correspond to the Swift XRT spectra taken a day apart in August 2007.  The additional Swift spectra correspond to the May 2007 (red), July 2007 (green), and June 2009 (magenta) observations.
\label{fig-comparexray}}
\end{figure}

In Figure~\ref{fig-comparexray}, we plot the ratio of the data/model of the XMM-Newton pn and five Swift XRT spectra, using the best-fit model and parameters described in the previous section for the pn spectrum.  Clearly, there is variability in both the spectral shape and the flux between the observations.  In particular, we find that the two Swift XRT spectra from August 2007 are at a much lower flux level than the additional observations.  The December 2009 pn spectrum corresponds to the highest flux level.


Allowing for a change in flux between the observations with the addition of a {\tt constant} model, we find a large improvement in the fit with a reduced $\chi^2$ value of 1.46.  Fixing the flux constant relative to the pn observation, we find constant values for the Swift XRT observations of 0.50 (2007 May), 0.60 (2007 July), 0.18 (2007 August 19), 0.19 (2007 August 20), and 0.64 (2009 June).  We find that allowing an intrinsic neutral absorbing column density to vary does not improve the fit.  

For each of the X-ray spectra, we individually fit a blackbody + power-law model.  We fixed the temperature of the blackbody component to 0.10\,keV and the Galactic absorption to the \citet{1990ARAA..28..215D} value.  In place of the XSPEC {\tt powerlaw} model, we used the pegged power-law model ({\tt pegpwrlw}).  This model is essentially the same as the power-law model, but the normalization is the flux in a defined band (we used 0.3--10\,keV).  The advantage of this model is that it allows us to determine error-bars on the power-law flux.  The best-fit parameters for each of the X-ray spectra are found in Table~\ref{tbl-xrayspectra}.  The errors on the blackbody normalization are large for the Swift observations.  However, we find that the XMM-Newton pn spectrum has the highest measured blackbody flux normalization.  It is unclear whether the higher value in the pn (by a factor of ten over the XRT-measured values) is a result of calibration differences or a physical change in the emitted spectrum.

\begin{figure}
\centering
\includegraphics[width=14cm]{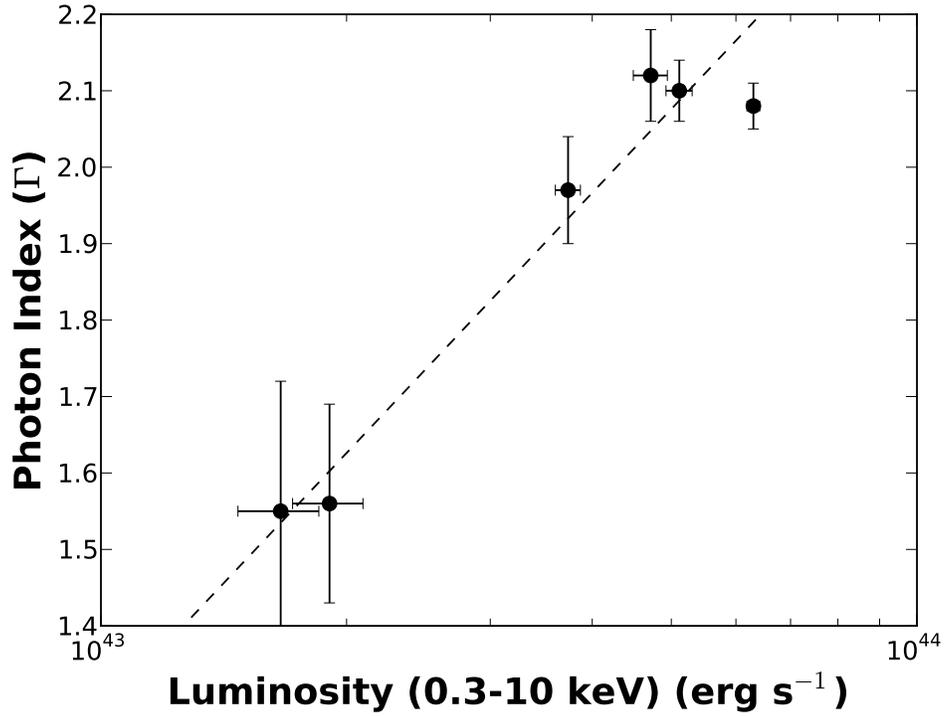}
\caption{The photon index versus 0.3--10\,keV luminosity for the X-ray observations of Mrk 817 is plotted, which span a range of approximately 2.5 years.  There is a clear correlation between the luminosity of the source and the photon index, with steeper photon indices corresponding to higher luminosities.  We plot the result of a linear regression fit, which shows a strong ($R^2 = 0.94$) correlation between $\Gamma$ and $L_X$.
\label{fig-gammalum}}
\end{figure}

The clearest result of our analysis is that the photon index is strongly correlated with the luminosity of the source, as shown in Figure~\ref{fig-gammalum}.  Using ordinary least-squares linear regression, we find that $\Gamma = (1.13 \pm 0.10) \log L_{0.3-10\,{\rm keV}} + (-47.33 \pm 4.38)$, with $R^2 = 0.94$ or Spearman correlation coefficient of $r = 0.77$, corresponding to $P = 0.05$.  Correlations between $\Gamma$ and the absorption-corrected 2--10\,keV luminosity are perhaps seen in high redshift samples (e.g.,~\citealt{2004ApJ...605...45D,2008AJ....135.1505S}), but not in low redshift ($z \le 0.1$) samples (e.g.,~\citealt{2000ApJ...531...52G,2009ApJ...690.1322W}).  While low redshift samples do not show the correlation, a correlation exists through measurements from multiple observations of individual low redshift sources (e.g.,~\citealt{1993ARAA..31..717M,2008ApJ...674..686W}).  The results from our analysis of the Mrk 817 X-ray spectra are consistent with previous results, where the photon index becomes steeper for higher luminosities and therefore higher accretion rates.

While broad-band X-ray spectra of Mrk 817 are not available over as long of a time period as in the UV, Mrk 817 was observed twice in the soft X-rays (0.1--2.4\,keV) with ROSAT \citep{1988ApOpt..27.1404A,1987SPIE..733..519P} in the early 1990s.  \citet{1996ApJ...471..190R} find a PSPC count rate of $0.3992 \pm 0.0095$\,ct\,s$^{-1}$ from a spectrum from a 4679\,s exposure (Dataset ID: RP701632N00) taken 1993-11-06.  Additionally, Mrk 817 was included in the ROSAT All-Sky Survey bright source catalog \citep{1999A&A...349..389V}, with a PSPC count rate of $0.102 \pm 0.017$\,ct\,s$^{-1}$ from a 542\,s exposure taken 1990-12-06.  

Using canned ROSAT PSPC response files and an assumed 0.1--2.4\,keV spectrum of a $\Gamma = 2.0$ power-law absorbed by the  \citet{1990ARAA..28..215D} value towards Mrk 817, we estimate a conversion between the ROSAT count rates and flux ($8 \times 10^{-12}$\flux at 1\,ct\,s$^{-1}$).  With this conversion, we find that the luminosities in the ROSAT observations are $(6.03 \pm 0.14) \times 10^{42}$\,ergs\,s$^{-1}$ (1993 observation) and  $(1.54 \pm 0.25) \times 10^{42}$\,ergs\,s$^{-1}$ (1990 observation).  To compare the luminosity of Mrk 817 over the nearly 20\,yr time span covered from the ROSAT through XMM-Newton observations, we calculated the 0.1--2.4\,keV luminosity for the Swift XRT and XMM-Newton pn observations using the best-fit models in Table~\ref{tbl-xrayspectra}.  

\begin{figure}
\centering
\includegraphics[width=14cm]{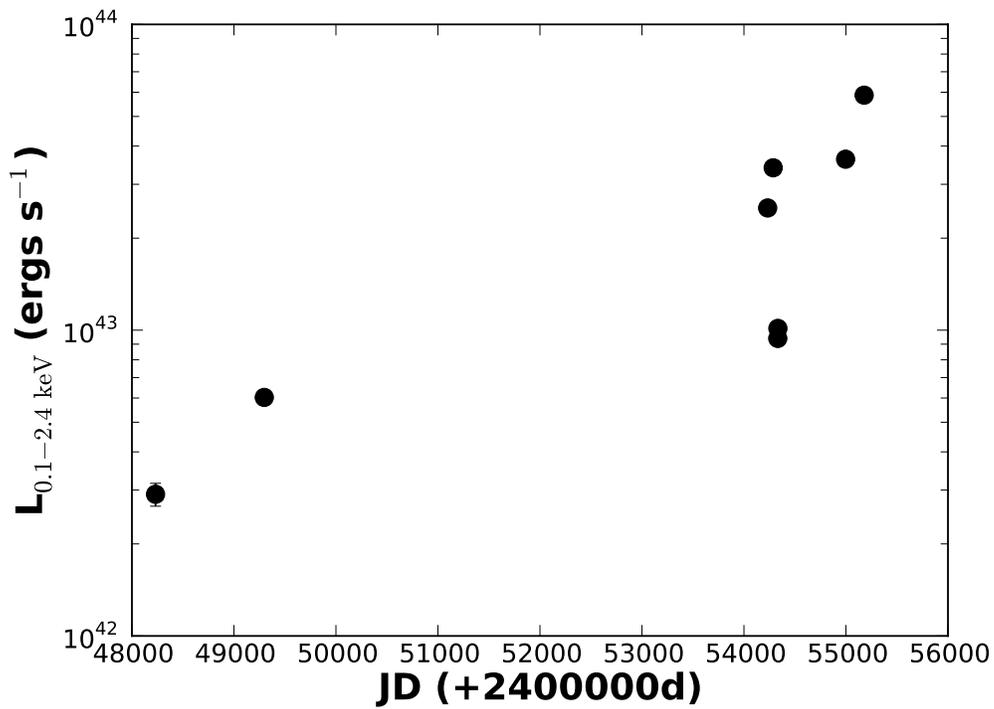}
\caption{The X-ray light curve of Mrk 817 is plotted, including the 0.1--2.4\,keV luminosity measured from two ROSAT PSPC observations, five Swift XRT observations, and one XMM-Newton observation.  The X-ray luminosity of Mrk 817 is clearly variable, with the highest luminosity observation 39 times that of the lowest luminosity observation.
\label{fig-xray_lc}}
\end{figure}

In Figure~\ref{fig-xray_lc}, we plot a light curve of the observed X-ray luminosity in the 0.1--2.4\,keV band constructed from the ROSAT, Swift XRT, and XMM-Newton observations of Mrk 817.  Variability is clear, with the 1990 ROSAT observation being 39 times less luminous than the 2009 XMM-Newton observation.  A factor of four variability is seen between the two ROSAT observations, which are separated by three years.  Additionally, in the Swift XRT observations, which also span approximately three years, there is a factor of four variability.  Thus, the X-ray spectra of Mrk 817, which span $\sim 20$\,years, show a higher level of variability than was observed in the UV spectra (from IUE, GHRS, and COS).

\subsubsection{Variability from the simultaneous Optical/UV/X-ray data}
To test whether the variability in the X-ray band is correlated with optical/UV variability in Mrk 817, we utilized XMM-Newton Optical Monitor (OM; \citealt{2001A&A...365L..36M}) and Swift Ultraviolet Optical Telescope (UVOT\footnote{See the UVOT website at: \url{http://www.swift.psu.edu/uvot/}}) data that were taken simultaneously with the X-ray spectra.  To extract the OM fluxes, we ran the standard SAS {\tt omichain} script to produce processed OM images.  The fluxes were obtained from the {\tt omdetect} algorithm, which extracts the flux for point-like sources in the images.  OM images were taken in both the UVW1 ($\lambda_c = 2910$\,\AA) and UVM2 ($\lambda_c = 2310$\,\AA) filters for our XMM-Newton observation.  UVOT observations were available for three of the five Swift observations of Mrk 817 (00036620001, 00036620002, and 00037592001).  To determine the flux of our source in the Swift UVOT observations, we ran the Swift task {\tt uvotsource} on the processed UVOT images, downloaded from HEASARC.  We used a source region centered on Mrk 817 with a 5$\arcsec$ circular aperture.  A background region was selected on the image in an area free from bright sources.  In Table~\ref{tbl-uvot}, we list the measured fluxes from each of the available filters.  Since Mrk 817 is a nearby face-on spiral, the extracted OM/UVOT fluxes are a combination of both the AGN and stellar bulge emission.

\begin{figure}
\centering
\includegraphics[width=14cm]{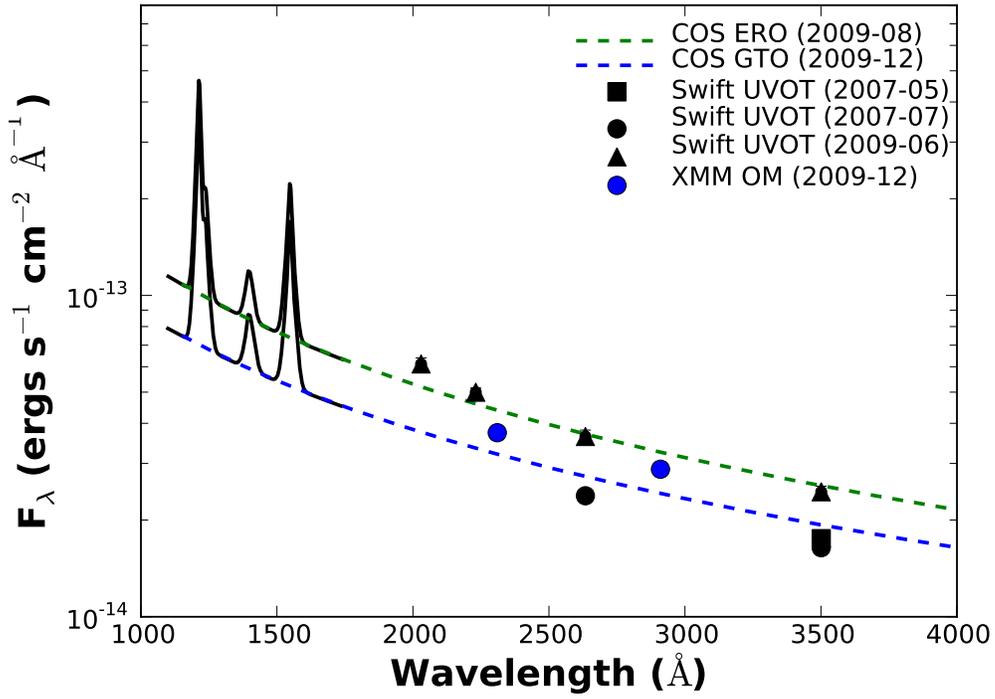}
\caption{Plotted are the UV-optical fluxes obtained from the Swift UVOT and XMM-Newton OM observations that were simultaneous to the X-ray observations.  Additionally, the best-fit emission line (solid lines, shown in the rest wavelength of Mrk 817) and continuum fits from our analyses of the COS spectra are shown (dashed lines), extended into the optical band.  The UVOT and OM fluxes are at a similar level to the values predicted by the COS observations, particularly for the OM observations which were taken within two weeks of the COS GTO spectra and are within 10\% of the values predicted by the COS GTO continuum.
\label{fig-sed}}
\end{figure}

In Figure~\ref{fig-sed}, we plot the observed fluxes obtained from UVOT and OM U, UVW1, UVM2, and UVW2 filters.  We also plot the continuum values measured from our COS observations and extended to 4000\,\AA.  We find that the UVOT and OM fluxes are well-matched to the slope and flux of the COS observations.  In particular, the OM values, taken within two weeks of the COS GTO observation, are $\la 10$\% higher than the estimated continuum from COS.  As shown in the figure, the scale of optical/UV variability over the Swift, COS, and XMM-Newton observations is minimal ($\sim 30$\%) over the 2.5\,year time frame.  Comparing the Swift UVOT U-band fluxes (the only filter for which we have data for the three Swift observations), we find that the highest flux observation is $(33 \pm 6)$\% higher than the lowest flux observation.  The X-ray variability is comparable, with a $(27 \pm 5)$\% change between the highest and lowest flux observation.  However, the observation with the highest X-ray flux does not correspond to the observation with the highest U-band flux.  Therefore, there is no obvious correlation between the X-ray and U-band flux.

The availability of simultaneous, multi-band data allows the construction of multi-epoch SEDs for Mrk 817 along with simple model fits to the data.  This illustrates how the value of $\alpha_{\rm OX}$, which is the optical-to-X-ray spectral slope, or X-ray loudness, defined as $\alpha_{\rm OX} = -0.384 \log(F_{2 \rm keV}/F_{2500 \rm \AA})$ \citep{1979ApJ...234L...9T}, varies between different observations.  We follow the approach outlined in \citet{2009MNRAS.399.1553V} for generating the SEDs, using the \textsc{ftool} \textsc{flx2xsp} to convert the XMM-Newton OM and Swift UVOT photometry into data files suitable for fitting in \textsc{xspec}.  We constructed SEDs for the three Swift observations with IDs 00036620001, 00036620002 and 00037592001 and the new XMM-Newton observation.  A simple disk plus absorbed power-law model was then fit to the optical/UV through X-ray data points (full model used: {\tt diskpn} + {\tt wabs}({\tt zwabs}({\tt bknpower}))), with the model parameters constrained as specified in \citealt{2009MNRAS.399.1553V}).  The normalization of the {\tt diskpn} component was constrained using the reverberation mass estimate from  \citet{2009ApJ...697..160B} and Galactic neutral hydrogen was included via the {\tt wabs} component.  

\begin{figure}
\centering
\includegraphics[width=14cm]{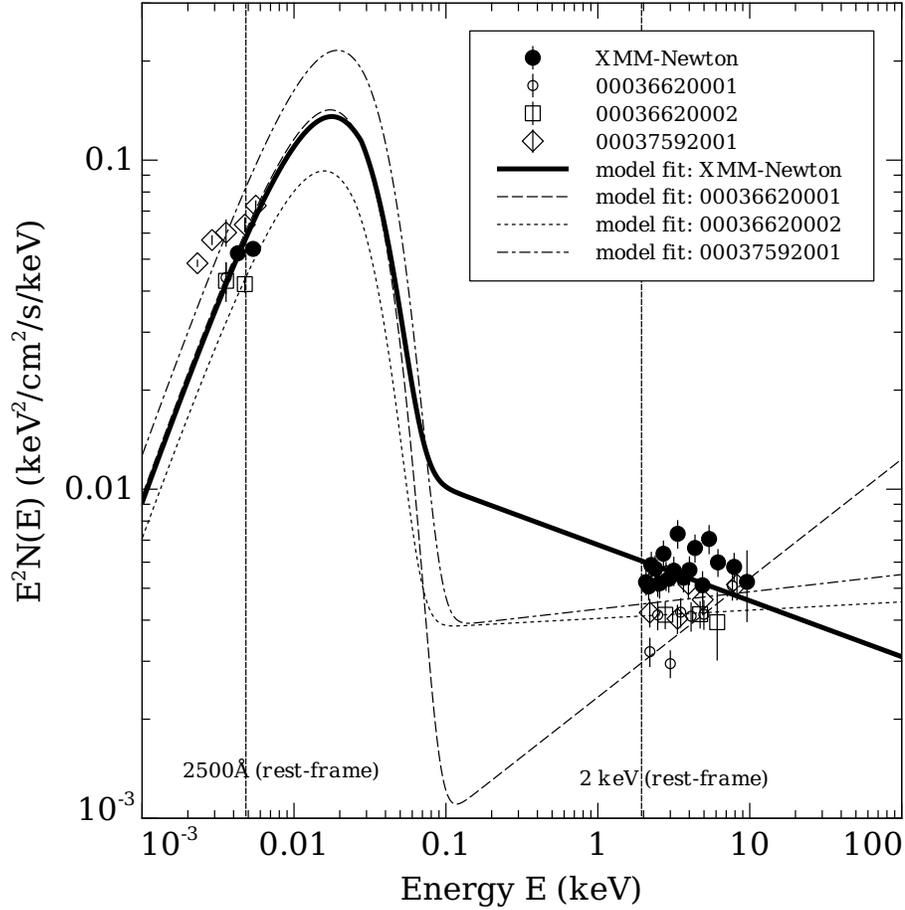}
\caption{Plotted are the spectral energy distributions from XMM-Newton and Swift observations with associated optical/UV data from OM/UVOT.  A model consisting of an accretion disk model with an absorbed power-law component was fit to each of the individual observations.  The optical/UV portion of the SED does not change significantly, while there is a noticeable difference in the X-ray portion. 
\label{fig-realsed}}
\end{figure}

The full SEDs for the four observations under discussion, along with their respective model fits (with any absorption removed to show the intrinsic SED model), are shown in Figure~\ref{fig-realsed}.  The model fits were then used to determine the value of $\alpha_{\rm OX}$, by computing the monochromatic luminosities at rest-frame 2500\,\AA\,and 2\,keV.  We obtain the following values of $\alpha_{\rm OX}$ for the observations: XMM-Newton: $-1.353^{+0.009}_{-0.011}$, Swift 00037592001: $-1.51^{+0.08}_{-0.16}$ and Swift 00036620001: $-1.51^{+0.05}_{-0.07}$.  The computed $\alpha_{\rm OX}$ value for the observation Swift 00036620002 is omitted due to its high uncertainty, since UVOT data was available in only one filter (U).  The XMM-Newton observation has the highest X-ray luminosity ($5.9 \times 10^{43}$\,ergs\,s$^{-1}$ in the 0.3--10\,keV band) and the smallest value of $\alpha_{OX}$.  The two Swift observations with multiple UVOT filter observations are consistent with having the same $\alpha_{OX}$ values.  The relationship between $\alpha_{OX}$-L$_X$ is not entirely clear, however, since the Swift observations have the same $\alpha_{OX}$ value but vary in luminosity by $26.9 \pm 4.6$\%.  Additionally, the XMM-Newton observation is only $18.9 \pm 4.3$\% more luminous than the Swift 00037592001 observation, but has a significantly different $\alpha_{OX}$ value.



\section{Conclusions}\label{sect-conclusion}
In this paper, we present an analysis of several UV and X-ray spectra of the Seyfert 1.5 Mrk 817.  The UV spectral analysis includes new high-resolution spectra from the Cosmic Origins Spectrograph, as well as archived spectra from IUE and HST GHRS.  These observations span nearly 30 years and allow us to search for spectral variability on decade time scales.  The X-ray spectral analysis includes a spectrum from XMM-Newton, taken within two weeks of the second COS observation, as well as five lower signal-to-noise ratio spectra with Swift XRT.  These observations allow us to probe X-ray spectral variability on the 2.5\,year time scale covered.  In addition, luminosities from ROSAT allow us to determine variability in the soft X-ray band over a 20\,year period.  Finally, an analysis of the OM/UVOT observations, which were simultaneous with the X-ray observations, allows us to determine variability in the UV/X-ray SED.  In this section, we summarize and discuss the results of our analysis.

\subsection{Spectral Shape and Fundamental Properties}
The HST COS observations of Mrk 817 exhibit the broad UV emission lines (Ly$\alpha$, \ion{N}{5}, \ion{Si}{4} + \ion{O}{4}], and \ion{C}{4}) characteristic of an optical broad-line Seyfert.  Our analysis of the ERO and GTO observations, separated by $\sim 5$\,months, finds similar power-law slopes ($F_{\lambda} \propto \lambda^{-1.30}$) and FWHMs for the emission lines (e.g., we find a broad component of $\sim 3000$\km\,and a narrower component of $\sim 1000$\km\,for the \ion{C}{4} emission).  The \ion{C}{4} emission is blue-shifted by values similar to other AGNs (blue-shifted by 50--340\,\km), but there is no correlation between the shifts in the \ion{C}{4}/\ion{Si}{4}/\ion{N}{5}/Ly$\alpha$ emission components.  Further, these velocity shifts change between observations.

No strong intrinsic absorption lines (i.e., Ly$\alpha$, \ion{N}{5}, \ion{C}{4}) are detected in the COS observations.  However, four high-velocity cloud absorbers and four intrinsic Ly$\alpha$ absorption lines are detected in the GHRS and COS spectra (as well as four absorbers detected at low-significance, see \S~\ref{sect-absorbers}).  Corresponding features are not detected in the more ionized lines.

The X-ray spectra of Mrk 817 are well-characterized by a blackbody + power-law model.  The temperature of the blackbody component, used to characterize the soft excess, of $kT \approx 0.1$\,keV, is typical of Seyfert 1s (e.g., \citealt{2005AA...444...79M,2009ApJ...690.1322W}) and quasars \citep{2004MNRAS.349L...7G}.  However, the power-law slope is steeper than the average photon index of similar AGN.  We find $\Gamma \ga 2.0$ in all but the lowest flux observations; while typical values for local Seyfert 1s detected in the hard X-rays with Swift's BAT (of which, Mrk 817 is included in the 22-month Swift BAT catalog) are $\Gamma = 1.78 \pm 0.24$ \citep{2009ApJ...690.1322W}.  Other than the steep slope, there are no unusual features in the spectrum.  Additionally, we find no evidence for either neutral or ionized absorption.

Using the flux and continuum properties derived from our spectral analysis, we determine fundamental properties of Mrk 817 to determine how it compares to local Seyferts.  The X-ray 2--10\,keV luminosity, which we compute using the 2--10\,keV flux of $1.45 \times 10^{-11}$\,\flux~from the XMM-Newton spectrum, can be used to estimate the bolometric luminosity of the AGN.  Using a bolometric correction of 35, typical of unabsorbed AGN \citep{2005AJ....129..578B} and within the range of parameters determined from our SED fits to the simultaneous optical/UV/X-ray data from Swift and XMM-Newton, we estimate $L_{bol} = 9.58 \times 10^{44}$\,erg\,s$^{-1}$.  This value is consistent with local ($\langle{z}\rangle = 0.03$) broad-line Seyferts selected in the 14--195\,keV band with the Swift BAT detector, where the average bolometric luminosity is $7.4 \times 10^{44}$\,erg\,s$^{-1}$, with a range of values from $\sim 9 \times 10^{43}$ -- $6 \times 10^{45}$\,erg\,s$^{-1}$ \citep{2010ApJ...710..503W}.  The black hole mass of Mrk 817, $\log {M}/{M}_{\sun} = 7.69 \pm 0.07$ as derived from reverberation mapping \citep{2009ApJ...697..160B}, is also typical of local AGN \citep{2002ApJ...579..530W} and the BAT AGN sample in particular ($\langle \log {M}/{M}_{\sun}\rangle = 7.87 \pm 0.66$; \citealt{2010ApJ...710..503W}).  However, the estimated accretion rate ($L_{bol}/L_{Edd}$, where $L_{Edd} = ({M}/{M}_{\sun}) \times 1.3 \times 10^{38}$\,erg\,s$^{-1}$) of 0.14 is higher than the average accretion rate of 0.034 (with a range of values from $\sim$ 0.001--0.5) determined for broad-line Seyferts in the BAT sample \citep{2010ApJ...710..503W}.  Thus, Mrk 817 is in many ways (mass, luminosity, UV spectral shape) a typical broad-line Seyfert.  Since the accretion rate is believed to determine the hard X-ray spectral slope in AGNs, with high spectral slopes correlating with high accretion rates \citep{2006ApJ...646L..29S}, it is not surprising that Mrk 817 has both a slightly higher photon index and accretion rate than the average local Seyfert 1. 

\subsection{Intrinsic Absorption}

Four intrinsic absorption features were detected embedded in the broad Ly$\alpha$ emission of Mrk 817.  These features include a $-4250$\km, $-4100$\km, $-3550$\km, and $-2600$\km\,absorber.  The strength of these features vary between the 1997 GHRS and 2009 COS observations, with the ratio of EW between GHRS/COS being $5.7 \pm 1.3$, $2.2 \pm 1.1$, $3.9 \pm 3.0$, and $\sim 0.3$, respectively.  The width of the absorption features are consistent between the GHRS and COS observations. No intrinsic \ion{C}{4}~$\lambda\lambda$1548.20,1550.77 and \ion{N}{5}~$\lambda\lambda$1238.82,1242.80 absorption features are detected at the same velocities.  However, given the weakness of the intrinsic Ly$\alpha$ absorbers in the COS spectra (EW $\sim 35$\,m\AA), this absence is expected.
For similar ionic column densities, the ratio of optical depth between two lines should be proportional to the ratio of their $f\lambda$ values (where $f$ is the oscillator strength and $\lambda$ is the wavelength of the transition). Under these conditions, $\tau$(Ly$\alpha$)/$\tau$(\ion{C}{4} $\lambda$1548)=1.7 and $\tau$(Ly$\alpha$)/$\tau$(\ion{N}{5} $\lambda$1238)=2.6.
For a photo-ionized plasma with a fixed total hydrogen column density, the column density of each ion is a strong function of the ionization parameter $U$.  However, as shown in figure 2 of \citet{2007ApJ...658..829A},
for solar abundances, the column densities of \ion{C}{4} and \ion{N}{5} are always smaller than that of \ion{H}{1}, irrespective of the value of $U$.  Combined with the stronger expected optical depth of the Ly$\alpha$ line,  a weak absorption feature of the latter should not show related features in \ion{C}{4} and \ion{N}{5}.  The same figure shows that under similar conditions \ion{O}{6} absorption is  expected for $\log{U}\sim-0.5$, where for solar abundances the \ion{O}{6} column density is 8 times larger than that of \ion{H}{1},
while $\tau$(Ly$\alpha$)/$\tau$(\ion{O}{6} $\lambda$1032)=3.7 for the same column densities. Under these conditions, we expect $\tau$(\ion{O}{6} $\lambda$1032)
to be twice that of Ly$\alpha$.  This may explain why \ion{O}{6} and Ly$\beta$ features are detected in the same FUSE data (For the same column densities, $\tau$(Ly$\beta$)/$\tau$(\ion{O}{6} $\lambda$1032)=0.6.). The photo-ionization curves also show that $\tau$(\ion{O}{6} $\lambda$1032)/$\tau$(Ly$\alpha$) drops below 1 for $\log{U}>0.4$ and $\log{U}<-1$. Similarly, with such a weak Ly$\alpha$ absorption feature, we do not expect to detect the \ion{O}{7} and \ion{O}{8} absorption edges in the near-simultaneous X-ray data.



The strongest GHRS Ly$\alpha$ absorber at $-4250$\,\km, which is $\sim 5$ times weaker in the COS spectra, is likely associated with the strongest \ion{O}{6} absorber seen in the four FUSE observations from 2000--2001 \citep{2008AJ....136.1201D}.  As we found for the GHRS/COS observations, the \ion{O}{6} absorber, whose velocity changes slightly from $-4198$\km~to $-4144$\km~throughout the observations, also varies in EW (from 0.60\,\AA~to 0.33\,\AA).  Both the strength and velocity of the \ion{O}{6} absorber decreased over time -- however, this decrease was not correlated with changes in the Far-UV luminosity of Mrk 817.  Additional weak \ion{O}{6} absorption lines were also found to appear throughout the set of four FUSE observations, with no obvious correlation to the UV luminosity.  In a similar way, a comparison of the COS observations with the GHRS observation shows no correlation between UV luminosity (in the continuum or emission lines) and the equivalent width of the absorber.  


Mrk 817 is not alone in exhibiting variable absorbers.  Variability in UV absorption lines has been detected in several Seyfert 1s, including NGC 4151 (e.g.,~\citealt{1981MNRAS.196..857P,1985MNRAS.215....1B,2006ApJS..167..161K}), NGC 3516 (e.g.,~\citealt{1983ApJ...267..515U}), NGC 5548 (e.g.,~\citealt{1993ApJ...416..536S,2009ApJ...698..281C}), and NGC 3783 (e.g.,~\citealt{1996ApJ...465..733M,2005ApJ...631..741G}).  The cause of variability of intrinsic absorbers is attributed to a change in ionization state or shielding of the photo-ionizing continuum of an outflowing wind or bulk motions of an absorbing cloud/clouds.


A potential cause of the change in the absorbers in Mrk 817 is from bulk motions of clouds along our line-of-sight.  As pointed out in \citet{2008AJ....136.1201D}, weak absorption -- as in the \ion{O}{6} absorbers in Mrk 817 -- is seen in spectra where this is the case.  Further, the fact that there is little change in the UV luminosity between the GHRS and COS spectra, while the strength of the Ly$\alpha$ absorber decreases significantly, also favors this model.  This would suggest that ionization is not causing a change in the absorber.  Based on the assumption of an absorbing cloud moving radially out of the line of sight in the 12.5 years between the GHRS and COS observations, this implies that the cloud moved a distance about $10^{17}$\,cm ($\ll 1$\,pc). 

Based on the X-ray measured upper limits on the \ion{O}{7} and \ion{O}{8} absorption edges, we can estimate an upper limit on the column density of the absorbing cloud to compare with UV-derived measurements.  Using the cross-sections ($\sigma$) for \ion{O}{7} and \ion{O}{8} from 
\citet{1996ApJ...465..487V} and the relation that $\tau = \sigma {\rm N}_{\rm O, ion}$, we find upper limits on the column density as N$_{\rm O VII} \la 4.0 \times 10^{17}$\,cm$^{-2}$ and N$_{\rm O VIII} \la 4.7 \times 10^{17}$\,cm$^{-2}$.  The typical range of column densities in \ion{C}{4} is well below the limit we find for \ion{O}{7} and \ion{O}{8}, with N$_{\rm C IV} = (0.1$--$14) \times 10^{14}$\,cm$^{-2}$ \citep{1999ApJ...516..750C}.  This X-ray derived limit is also much higher than the estimated \ion{H}{1} column from the Ly$\alpha$ absorption line ($\sim 4.3 \times 10^{13}$\,cm$^{-2}$ in the GHRS and $\sim 7.1 \times 10^{12}$\,cm$^{-2}$ in COS, assuming a fully covering absorber).  Therefore, it is possible that an outflow is present, but it has a column density too low to be detected in the CCD spectra analyzed.

\subsection{Variability}

With available observations of Mrk 817 from multiple observatories spanning decades, we were able to probe variability in flux and spectral shape across the UV and X-ray bands.  In the UV, from analysis of the IUE, GHRS, and COS spectra, we do not find drastic changes in either the line or continuum flux of Mrk 817.  As with other AGN, the continuum flux is correlated with the emission line flux of \ion{C}{4} in accordance with the Baldwin effect \citep{1977ApJ...214..679B,1994ApJ...436..678O}.  Measured changes in UV flux are limited to a factor of $\la 2.3$ between the minimum and maximum measured values over the 30\,year time-span probed.  Additionally, in support of our spectral analysis, we find that the UV photometry from XMM-Newton OM and Swift UVOT observations are at a similar level as the UV continuum measured from the COS spectra.  

Variability in the X-ray spectra is much more pronounced than in the UV.  In particular, we find that the X-ray spectrum of Mrk 817 varies in both luminosity and spectral shape.  The photon index measured from a power-law component varies from $\Gamma \approx 1.5$--$2.1$ between the five Swift XRT and the XMM-Newton spectra, which span 2.5\,years.  The variability in the spectral slope is correlated strongly with the observed X-ray luminosity, where the steepest slopes are observed at the highest X-ray luminosities.  

Two ROSAT flux measurements from the early 1990s have allowed us to search for long-term variability in the soft X-ray band luminosities from the ROSAT, Swift, and XMM-Newton observations.  We find that the X-ray luminosity of Mrk 817 changes markedly.  Shorter term variability is observed in the Swift XRT observations, with a factor of four change in variability over $\sim 2$\,years.  On the long term, we find that the 1990 ROSAT luminosity is $\sim 40$ times smaller than the luminosity measured from the 2009 XMM-Newton observation.    The factor of 40 variability seen over 20 years in the X-ray observations contrasts with the factor of 2.3 seen in the UV spectra.  Thus, Mrk 817 is clearly more variable in the X-ray than UV band.      

To search for a possible relationship between the UV and X-ray luminosities, we analyzed simultaneous optical/UV/X-ray data from XMM-Newton and Swift.  We found that the optical/UV fluxes were at similar levels in the four observations with OM/UVOT photometry.  The optical-X-ray SED of Mrk 817 changes between observations, as indicated by $\alpha_{OX}$.  However, these changes are dominated by the change in X-ray luminosity/spectral shape.  Therefore, it appears that the UV is not correlated with the X-ray and that even in our simultaneous UV/X-ray data, we find more variability in the X-ray than the UV.  Unfortunately, however, there is no UVOT data for the two Swift observations with the lowest X-ray luminosities, making it impossible to determine whether the UV luminosity is low when the X-ray luminosity is at a much lower level than sampled with the available simultaneous data.

Finally, our analysis of the UV and X-ray observations suggests that more data are needed before we can rule out the possibility that the Ly$\alpha$ absorbers are the result of an ionized outflow.  While we did not find an X-ray \ion{O}{7} or \ion{O}{8} absorption edge strongly detected in our data, it is possible that a high signal-to-noise grating observation could reveal ionized X-ray absorption/emission at a lower column density than we could probe with the CCD spectra.  Additionally, while the UV luminosity does not change significantly, it is possible that the X-ray spectrum during the 1997 GHRS observation was significantly different than in 2009, when the COS spectrum was taken.  Further, our results of both a lack of a correlation between the UV and X-ray luminosities and a nearly constant UV luminosity over 30\,years while the X-ray luminosity changes by an order of magnitude, coupled with the possibility that changes in ionization of an outflow are the result of changes in the X-ray emission, suggest that the factor of 5 change in the Ly$\alpha$ absorption is potentially explained by a change in the X-ray spectrum.  Therefore, future multi-wavelength observations, with COS and X-ray grating data from Chandra/XMM-Newton, of Mrk 817 are vital to unveil both the true nature of the absorbers and any potential connection between the UV and X-ray emission.  



\acknowledgements
The authors acknowledge support through NASA grant HST-NNX08AC14G, in support of the COS GTO program (PI Green).  LMW acknowledges support through NASA grant HST-HF-51263.01-A, through a Hubble Fellowship from the Space Telescope Science Institute, which is operated by the Association of Universities for Research in Astronomy, Incorporated, under NASA contract NAS5-26555.  WNB and RVV acknowledge support from NASA grant NNX09AQ02G and NASA ADP grant NNX10AC99G.
Some of the data presented in this paper were obtained from the Multimission Archive at the Space Telescope Science Institute (MAST). STScI is operated by the Association of Universities for Research in Astronomy, Inc., under NASA contract NAS5-26555. Support for MAST for non-HST data is provided by the NASA Office of Space Science via grant NNX09AF08G and by other grants and contracts.  This work utilizes observations obtained with XMM-Newton, an ESA science mission with instruments and contributions directly funded by ESA Member States and NASA.

{\it Facilities:} \facility{HST()}, \facility{Swift()}, \facility{XMM()}, \facility{IUE()}

\bibliography{MyBibtex.bib}

\begin{deluxetable}{llccc}
\tablecaption{Summary of Ultraviolet Observations\label{tbl-UV}}
\tablewidth{0pt}
\tablehead{
\colhead{Observation} & \colhead{Date} & \colhead{Exp Time (ks)} & \colhead{Spectral Coverage}
& \colhead{$\lambda/\Delta\lambda$\tablenotemark{\dagger}}
}
\startdata
COS ERO & 2009-08-04 & 3.4 & 1135--1795\,\AA & 18000\\
COS GTO & 2009-12-28 & 3.0 & 1135--1795\,\AA & 18000\\
\hline 
GHRS & 1997-01-12 & 25.0 & 1223--1259\,\AA & 18000\\
\hline 
IUE SWP15447 & 1981-11-07 & 16.2 & 1120--2010\,\AA & 200\\
IUE SWP17442 & 1982-07-18 & 16.2 & 1120--2010\,\AA & 200\\
\enddata
\tablenotetext{\dagger}{Approximate spectral resolution at 1200\AA.}
\end{deluxetable}

\begin{deluxetable}{lll}
\tablecaption{Power law Best-fit Parameters\label{tbl-powerlaw}}
\tablewidth{0pt}
\tablehead{
\colhead{Parameter} & \colhead{COS ERO} & \colhead{COS GTO}
}
\startdata
$a_0$ & $0.63 \pm 1.02$ & $0.45 \pm 1.02$\\
$\alpha$ & $1.30 \pm 0.05$ & $1.22 \pm 0.07$ \\
\enddata
\vspace{0.2cm}

Parameters of a simple power law fit, of the form $a_0 \times (\lambda/1750$ \AA$)^{-\alpha}$, where $a_0$ is the continuum flux density at 1750\,\AA~(in units of $10^{-13}$ \flux\,\AA$^{-1}$) and $\alpha$ is the power law index.
\end{deluxetable}

\begin{deluxetable}{ccccc}
\tabletypesize{\scriptsize}
\tablecaption{Emission Line Best-fit Parameters for the COS ERO spectrum\label{tbl-emission-ero}}
\tablewidth{0pt}
\tablehead{
     \colhead{Component} & 
     \colhead{v${\rm (off)}$ (\km)} & 
     \colhead{FWHM (\km)} & 
     \colhead{$EW$ (\AA)} & 
     \colhead{Log Flux}
}
\startdata
\multicolumn{5}{c}{\bf C\,IV 1550.774\,\AA}\\
\multicolumn{5}{c}{$\chi^2/{\rm dof} = 9402.9/9491$} \\
1 & $-337.6 \pm 15.1$ & $3145 \pm 31$ & $15.97 \pm 0.14$ & $-11.93 \pm 0.01$ \\
2 & $-126.1 \pm 15.4$ & $1051 \pm 30$ & $4.04 \pm 0.17$ & $-12.53 \pm 0.02$ \\
\hline

\multicolumn{5}{c}{\bf Si\,IV 1402.77\,\AA} \\
\multicolumn{5}{c}{$\chi^2/{\rm dof} = 21104.7/11489$} \\
1 & $381.3 \pm 22.6$ & \nodata & $4.45 \pm 0.04$ & $-12.43 \pm 0.01$ \\
2 & $432.2 \pm 55.5$ & \nodata & $0.45 \pm 0.02$ & $-13.15 \pm 0.01$ \\
\hline

\multicolumn{5}{c}{\bf N\,V 1242.80\,\AA}\\
\multicolumn{5}{c}{$\chi^2/{\rm dof} = 20224.6/7807$} \\
1 & $27.2 \pm 8.6$ & \nodata & $7.69 \pm 0.04$ & $-12.12 \pm 0.01$ \\
2 & $-100.0 \pm 2.9$ & \nodata & $1.32 \pm 0.02$ & $-12.89 \pm 0.01$ \\
\hline
\multicolumn{5}{c}{\bf Ly$\alpha$ 1215.67\,\AA}  \\
1 &  $-905.0 \pm 9.8$ & \nodata & $48.52 \pm 0.47$ & $-11.31 \pm 0.01$ \\
2 & $800.0 \pm 10.0$ & \nodata &  $12.82 \pm 0.17$ & $-11.89 \pm 0.01$ \\
3 & $-556.4 \pm 22.0$ & $1900 \pm 26$ & $17.20 \pm 0.38$ & $-11.76 \pm 0.01$ \\
\enddata
\vspace{0.2cm}

The results of multiple Gaussian fits, to the indicated UV emission lines, are shown.  The FWHM for Si\,IV, N\,V, and the first two components of Ly$\alpha$ are fixed to the values from the C\,IV fits.  The logarithm of the integrated line flux is recorded for each component (with units of ergs\,s$^{-1}$\,cm$^{-2}$).  We fit the doublets (C\,IV, Si\,IV, and N\,V) by fixing the width between the two lines of the doublet to the known separation of the lines.  The flux and FWHM of each component in the doublet were tied together, with the flux of the blue line in the doublet fixed to be twice that of the red line (the ratio of the oscillator strengths of blue/red is $\sim 2:1$ for each of these doublets).  The total flux for each doublet, then, is three times the total flux for the red line listed in the Table.
\end{deluxetable}

\begin{deluxetable}{ccccc}
\tabletypesize{\scriptsize}
\tablecolumns{5} 
\tablecaption{Emission Line Best-fit Parameters for the COS GTO spectrum\label{tbl-emission-gto}}
\tablewidth{0pt}
\tablehead{
     \colhead{Component} & 
     \colhead{v${\rm (off)}$ (\km)} & 
     \colhead{FWHM (\km)} & 
     \colhead{$EW$ (\AA)} & 
     \colhead{Log Flux}  
}
\startdata
\multicolumn{5}{c}{\bf C\,IV 1550.774\,\AA} \\ 
\multicolumn{5}{c}{$\chi^2/{\rm dof} = 8972.2/9491$} \\
1 & $-53.5 \pm 17.5$ & $3481 \pm 33$ & $18.05 \pm 0.14$ & $-12.03 \pm 0.01$ \\
2 & $-270.0 \pm 12.4$ & $1030 \pm 19$ & $5.00 \pm 0.14$ & $-12.59 \pm 0.01$ \\
\hline

\multicolumn{5}{c}{\bf Si\,IV 1402.77\,\AA} \\
\multicolumn{5}{c}{$\chi^2/{\rm dof} = 14262.0/11491$} \\
1 & $765.8 \pm 27.5$ & \nodata & $5.38 \pm 0.05$ & $-12.50 \pm 0.01$ \\
2 & $232.2 \pm 48.0$ & \nodata & $0.65 \pm 0.02$ & $-13.41 \pm 0.01$ \\
\hline

\multicolumn{5}{c}{\bf N\,V 1242.80\,\AA} \\ 
\multicolumn{5}{c}{$\chi^2/{\rm dof} = 11948.1/7807$} \\
1 & $-216.1 \pm 14.6$ & \nodata & $9.67 \pm 0.08$ & $-12.18 \pm 0.01$ \\
2 & $-100.0 \pm 4.9$ & \nodata & $1.81 \pm 0.04$ & $-12.91 \pm 0.01$ \\
\hline

\multicolumn{5}{c}{\bf Ly$\alpha$ 1215.67\,\AA}  \\
1 &  $-1154.2 \pm 8.8$ & \nodata & $57.57 \pm 0.30$ & $-11.40 \pm 0.01$ \\
2 & $800 \pm 5.8$ & \nodata &  $18.07 \pm 0.37$ & $-11.90 \pm 0.01$ \\
3 & $-56.7 \pm 28.1$ & $1833 \pm 34$ & $16.90 \pm 0.39$ & $-11.93 \pm 0.01$\\
\enddata
\vspace{0.2cm}

The notes to this table are the same as for Table~\ref{tbl-emission-ero}.
\end{deluxetable}

\begin{deluxetable}{lllll}
\tablecaption{Intrinsic Ly$\alpha$ Absorber Measurements\label{tbl-absorber}}
\tablewidth{0pt}
\tablehead{
\colhead{Observation} & \colhead{v(sys) (\km)} & \colhead{FWHM (\km)} & \colhead{EW (m\AA)} &
\colhead{$\log {\rm N_{\rm H I}}$ (cm$^{-2}$)}
}
\startdata
GHRS & $-4269 \pm 3$ & $125 \pm 8$ & $199 \pm 10$ & $13.63 \pm 0.02$ \\
COS ERO & $-4332 \pm 20$ & $113 \pm 17$ & $39 \pm 10$ & $12.9 \pm 0.1$ \\
COS GTO & $-4322 \pm 20$ & $< 133$ & $53 \pm 10$ & $13.0 \pm 0.1$ \\
Combined COS & $-4288 \pm 10$ & $110 \pm 17$ & $37 \pm 5$ & $12.85 \pm 0.08$ \\
\enddata
\vspace{0.2cm}

The parameters of Voigt profile fits to the strongest absorption feature in the GHRS and COS spectra include the systemic velocity of the absorber, the full-width at half maximum, equivalent width, and \ion{H}{1} column density (assuming a fully covering absorber).
\end{deluxetable}

\begin{deluxetable}{lcccccc}
\tabletypesize{\scriptsize}
\tablecolumns{7} 
\tablewidth{0pt} 
\tablecaption{Intrinsic Ly$\alpha$ Absorbers\label{tbl-intrinsic}}
\tablehead{
           \colhead{$\lambda_{\rm obs}$}     &
           \colhead{$c\,z_{\rm AGN}$\tablenotemark{a}} & 
           \colhead{GHRS $EW$}                &
           \colhead{GHRS FWHM}               &
           \colhead{COS $EW$}                 &
           \colhead{COS FWHM}                &
           \colhead{Absorber}                \\      
           \colhead{(\AA)}                   &
           \colhead{(\km)}                   & 
           \colhead{(m\AA)}                  &
           \colhead{(\km)}                   &
           \colhead{(m\AA)}                  &
           \colhead{(\km)}                   &
           \colhead{Notes}                   }      

\startdata
 1223.5\tablenotemark{b} & $-7410$ & $ 19\pm10$ & $ 43\pm23$& $ <6    $ & \nodata  & low-significance\tablenotemark{\dagger}  \\
 1224.2\tablenotemark{b} & $-7250$ & $130\pm10$ & $ 63\pm13$& $150\pm6$ & $ 75\pm8 $ & consistent     \\
 1234.7\tablenotemark{b} & $-4660$ & $ 33\pm15$ & $ 83\pm17$& $ <5    $ & \nodata  & low-significance\tablenotemark{\dagger}\\ 
 1236.3\tablenotemark{b} & $-4250$ & $200\pm10$ & $125\pm8 $& $ 35\pm5$ & $110\pm8 $ & variable \\
 1236.9\tablenotemark{b} & $-4100$ & $ 28\pm5 $ & $ 40\pm8 $& $ 13\pm3$ & $ 55\pm20$ & weaker component \\
 1239.2\tablenotemark{b} & $-3550$ & $ 54\pm8 $ & $122\pm17$& $ 14\pm5$ & $ 60\pm13$ & Galactic N\,V blend \\
 1241.0\tablenotemark{b} & $-3090$ & $ 34\pm5 $ & $ 40\pm10$& $ 34\pm3$ & $ 47\pm10$ & consistent \\
 1243.0                  & $-2600$ & $<5      $ & \nodata & $ 20:   $ & $58:     $ & Galactic N\,V blend \\
 1245.4\tablenotemark{b} & $-2010$ & $ 18:    $ & $ 75:    $& $ <5    $ & \nodata  & low significance\tablenotemark{\dagger}  \\
 1247.3\tablenotemark{b} & $-1540$ & $ 30\pm6 $ & $107\pm17$& $ <5    $ & \nodata  & low significance\tablenotemark{\dagger} \\
 1249.6                  & $ -980$ & $ 66\pm5 $ & $ 85\pm8 $& $ 64\pm5$ & $ 73\pm8 $ & consistent \\
 1251.8                  & $ -440$ & $ 14\pm4 $ & $ 25\pm8 $& $ 8\pm5 $ & $ 25:    $ & weak but consistent \\
\enddata
\tablenotetext{a}{Offset velocity from assumed AGN velocity of 9341~\km.}
\tablenotetext{b}{Absorber listed in Penton et al. (2000).}
\tablenotetext{\dagger}{~Low significance is defined as significance $< 10\sigma$.}
\end{deluxetable}

\begin{deluxetable}{lll}
\tablecaption{Best-fit Parameters for the IUE Spectra\label{tbl-iuespectra}}
\tablewidth{0pt}
\tablehead{
\colhead{Parameter} & \colhead{SWP15447} & \colhead{SWP17442}
}
\startdata
$a_0$\tablenotemark{\dagger} & $0.35 \pm 1.01$ & $0.41 \pm 1.01$\\
$\alpha$\tablenotemark{\dagger} & $0.55 \pm 0.07$ & $0.98 \pm 0.05$ \\
\ion{C}{4} doublet\tablenotemark{\ddagger} & $-11.55 \pm 0.04$ & $-11.72\pm0.08$\\
\ion{Si}{4} doublet\tablenotemark{\ddagger} &  $-12.20 \pm 0.71$& $-12.39 \pm $0.05\\
\ion{N}{5} doublet\tablenotemark{\ddagger} & $-11.79 \pm 0.04$ & $-12.21 \pm 0.22$\\
Ly$\alpha$\tablenotemark{\ddagger} & $-11.32 \pm 0.04$ & $-11.23 \pm 0.04$\\
\enddata
\tablenotetext{\dagger}{Parameters of a simple power law fit, of the form $a_0 \times (\lambda/1750$ \AA$)^{-\alpha}$, where $a_0$ is the continuum flux density at 1750\AA~(in units of $10^{-13}$ \flux\,\AA$^{-1}$) and $\alpha$ is the power law index. \\}
\tablenotetext{\ddagger}{The logarithm of the total line flux (ergs\,s$^{-1}$\,cm$^{-2}$) is shown for the indicated doublet or line.}
\end{deluxetable}

\begin{deluxetable}{llcc}
\tablecaption{Swift XRT Observations\label{tbl-xrt}}
\tablewidth{0pt}
\tablehead{
\colhead{Obs ID} & \colhead{UT Date} & \colhead{Exp Time} & \colhead{Count Rate\tablenotemark{\dagger}} \\
\colhead{} & \colhead{} & \colhead{(s)} & \colhead{(cts\,s$^{-1}$)}
}
\startdata
00036620001 & 2007-05-12 & 6670 & $0.56 \pm 0.09$\\
00036620002 & 2007-07-05 & 2355 & $0.70 \pm 0.02$ \\
00036620003 & 2007-09-19 & 2213 & $0.22 \pm 0.01$\\
00036620004 & 2007-09-20 & 2824 & $0.25 \pm 0.01$\\
00037592001 & 2009-06-14 & 5149 & $0.77 \pm 0.01$\\
\enddata
\tablenotetext{\dagger}{Count rate is given in the 0.3--10\,keV band.}
\end{deluxetable}

\begin{deluxetable}{lllll}
\tablecaption{Best-fit Parameters for the X-ray Observations\label{tbl-xrayspectra}}
\tablewidth{0pt}
\tablehead{
\colhead{Obs. Date} & \colhead{$A_{kT}$\tablenotemark{\dagger}} &
\colhead{$\Gamma$\tablenotemark{\dagger}} & \colhead{F(0.3--10\,keV)\tablenotemark{\dagger}} & \colhead{$\chi^2/dof$}
}
\startdata
2007-05-12 (XRT) & $1.37^{+1.46}_{-1.34}$ & $1.97^{+0.07}_{-0.05}$ & $1.98 \pm 0.07$ & 169.7/144\\
2007-07-05 (XRT) & $1.79$\tablenotemark{\ddagger} & $2.12^{+0.05}_{-0.06}$ & $2.50 \pm 0.12$ & 63.7/73 \\
2007-09-19 (XRT) & $2.32^{+0.99}_{-1.14}$ & $1.55^{+0.17}_{-0.16}$ & $0.88 \pm 0.10$ & 14.6/21 \\
2007-09-20 (XRT) & $2.00^{+0.87}_{-0.99}$ & $1.56^{+0.13}_{-0.12}$ & $1.01^{+0.10}_{-0.09}$ & 41.9/32 \\
2009-06-14 (XRT) & $1.95$\tablenotemark{\ddagger} & $2.10^{+0.03}_{-0.04}$ & $2.71^{+0.08}_{-0.10}$ & 131.3/150\\
2009-12-13 (pn) & $14.9^{+1.34}_{-1.41}$ & $2.08 \pm 0.03$ & $3.34 \pm 0.06$ & 391.9/339 \\
\enddata
\tablenotetext{\dagger}{Parameters are shown from a blackbody + power-law fit.  The blackbody temperature was fixed at 0.10\,keV and the flux normalization is recorded as $A_{kT}$ (in units of $10^{-5}$\,photons\,keV$^{-1}$\,cm$^{-2}$\,s$^{-1}$ at 1\,keV). Parameters of the power-law (using the {\tt pegpwrlw} model) include the photon index, $\Gamma$ where $F(E) \propto E^{-(\Gamma-1)}$, and F(0.3--10\,keV), the flux in the power-law component in the 0.3--10\,keV band in units of $10^{-11}$\,erg\,s$^{-1}$\,cm$^{-2}$.}
\tablenotetext{\ddagger}{Upper limit on the indicated parameter.}
\end{deluxetable}

\begin{deluxetable}{lllll}
\tablecaption{Measured Flux from the UVOT/OM Observations\label{tbl-uvot}}
\tablewidth{0pt}
\tablehead{
\colhead{Obs. Date} & \colhead{U\tablenotemark{\dagger}} & \colhead{UVW1\tablenotemark{\dagger}} & \colhead{UVM2\tablenotemark{\dagger}} &  \colhead{UVW2\tablenotemark{\dagger}} 
}
\startdata
2007-05-12 (UVOT) & $1.75 \pm 0.07$ & \nodata & \nodata & \nodata \\
2007-07-05 (UVOT) & $1.64 \pm 0.07$ & $2.38 \pm 0.11$ & \nodata & \nodata \\
2009-06-14 (UVOT) & $2.45 \pm 0.10$ & $3.64 \pm 0.17$ & $5.00 \pm 0.15$ & $6.17 \pm 0.27$ \\
2009-12-13 (OM) & \nodata & $2.88$ & $3.74$  & \nodata\\
\enddata
\tablenotetext{\dagger}{Fluxes are in units of $10^{-14}$\,erg\,s$^{-1}$\,cm$^{-2}$\,\AA$^{-1}$.  For the UVOT, the central wavelength of the filters corresponds to U at 3501\AA, UVW1 at 2634\AA, UVM2 at 2231\AA, and UVW2 at 2030\AA.  For the OM, the filters correspond to UVW1 at 2910\AA~and UVM2 at 2310\AA.}

\end{deluxetable}

\clearpage

\end{document}